\documentclass[11pt]{article}
\PassOptionsToPackage{table,dvipsnames}{xcolor}
\usepackage[final]{acl}
\usepackage{times}
\usepackage{amsmath}
\usepackage{amssymb}
\usepackage{latexsym}
\usepackage{booktabs}
\usepackage{float}
\usepackage{placeins}

\usepackage{array}
\usepackage{multirow}
\usepackage{makecell}
\usepackage{enumitem}
\usepackage[T1]{fontenc}
\usepackage[utf8]{inputenc}
\usepackage{microtype}
\usepackage{inconsolata}
\usepackage{graphicx}
\usepackage{xcolor}
\usepackage{tcolorbox}
\tcbuselibrary{breakable,skins}
\usepackage{caption}
\usepackage{pifont}
\usepackage[normalem]{ulem}
\usepackage{algorithm}
\usepackage{algpseudocode}
\usepackage{fontawesome5}
\usepackage{tikz}
\usetikzlibrary{positioning,calc,fit,backgrounds,shapes.misc,arrows.meta,shadows.blur}
\definecolor{userC}{RGB}{111,66,193}
\definecolor{userF}{RGB}{243,238,252}
\definecolor{sysC}{RGB}{230,126,34}
\definecolor{sysF}{RGB}{254,243,231}
\definecolor{badC}{RGB}{192,57,43}
\definecolor{goodC}{RGB}{39,130,76}
\definecolor{goodF}{RGB}{233,247,239}
\definecolor{retC}{RGB}{41,98,163}
\definecolor{retF}{RGB}{234,242,251}
\definecolor{revC}{RGB}{196,78,24}
\definecolor{revF}{RGB}{253,240,232}
\definecolor{navyC}{RGB}{31,59,115}
\definecolor{tealC}{RGB}{15,118,110}
\definecolor{tealF}{RGB}{230,244,242}
\definecolor{coralC}{RGB}{194,65,12}
\definecolor{coralF}{RGB}{253,238,230}
\definecolor{inkC}{RGB}{55,65,81}
\definecolor{slInk}{RGB}{30,41,59}
\definecolor{slMut}{RGB}{100,116,139}
\definecolor{slLine}{RGB}{203,213,225}
\definecolor{slBg}{RGB}{248,250,252}
\definecolor{slBub}{RGB}{241,245,249}
\definecolor{pBlue}{RGB}{37,99,235}
\definecolor{pBlueD}{RGB}{30,64,175}
\definecolor{pTeal}{RGB}{13,148,136}
\definecolor{pAmber}{RGB}{234,88,12}
\definecolor{pRed}{RGB}{220,38,38}
\definecolor{pGreen}{RGB}{22,163,74}
\definecolor{panelF}{RGB}{247,247,249}

\newcommand{\dataset}{\textsc{ConvRevise}}
\newcommand{\system}{\textsc{MemRevise}}
\newcommand{\cmark}{\ding{51}}
\newcommand{\xmark}{\ding{55}}

\title{\system{}: A Self-Revising Memory for Conversational Search}

\author{
  Mohammed Ali \quad Abdelrahman Abdallah \quad Adam Jatowt \\
  University of Innsbruck, Austria \\
  \texttt{\{mohammed.ali, abdelrahman.abdallah, adam.jatowt\}@uibk.ac.at}
}

\begin{document}
\maketitle

\begin{abstract}
In conversational search, users correct misunderstandings, revise their requests and return to earlier topics, so part of the history no longer reflects what they want. Existing benchmarks do not isolate such turns. Current search methods either re-read the history or keep a record of every turn, and none of them marks earlier content as dropped. We introduce \dataset{}, a benchmark of 4{,}429 test questions in six tiers of returns and revisions over TopiOCQA, CORAL and OR-QuAC. We also propose \system{}, which combines a revisable memory graph, a 3B writer distilled from teacher memories, and a search guided by the memory. The writer reads each turn once and adds a node that records its topic, its facts and the operations it performs, such as adding, abandoning or reactivating a topic. Topic status follows from these operations, so a dropped topic keeps its facts and can be restored. With a frozen retriever, \system{} reaches a mean MRR@10 of 0.701, against 0.589 for the best competitor, ConvSearch-R1-FT, and 0.617 for a writer distilled the same way without the memory, so the memory and the steps built on it add 0.084. \system{} is significantly better on TopiOCQA and OR-QuAC, statistically tied with prompted 32B rewriters on CORAL, and never significantly behind the strongest competitor of any tier.
\footnote{\dataset{}, our code and the trained writers will be released upon publication.}
\end{abstract}

\section{Introduction}
\label{sec:intro}

Conversational search enables users to navigate complex information needs through sequential, multi-turn dialogue. Because human interaction is highly contextual, users naturally omit established entities in follow-up queries, making context resolution a central focus of conversational information retrieval. Existing methodologies typically address this by rewriting the utterance into a standalone query \citep{canard,t5qr,chiq} or projecting the dialogue history into a unified dense representation \citep{convdr,chatretriever}. However, these approaches implicitly assume that conversational context accumulates as a purely additive process. In practice, search behaviour is highly dynamic: users frequently correct system misunderstandings, as studied in spoken dialogue \citep{hirschberg2001,litman2006}, revise their earlier requests, or return to previously explored topics. Consequently, after a correction or a revision, part of the established history no longer reflects what the user wants, and after a return, earlier context becomes relevant again. Since existing conversational search methods rely on full-history encoding or append-only memory, they lack an explicit way to mark which topics have been dropped or restored.

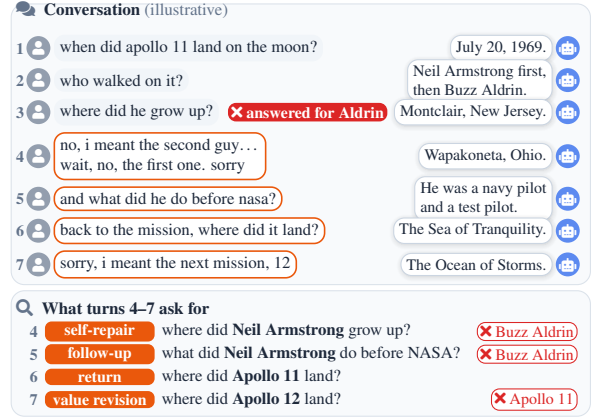
\begin{figure}[t]
\centering
\begin{tikzpicture}[
  font=\fontsize{6.3}{7.5}\selectfont, text=slInk,
  sh/.style={blur shadow={shadow blur steps=6, shadow xshift=0.25pt, shadow yshift=-0.5pt, shadow blur extra rounding=1pt, shadow opacity=18}},
  ubub/.style={rectangle, rounded corners=4pt, fill=slBub, inner sep=2.2pt, align=left},
  abub/.style={rectangle, rounded corners=4pt, fill=white, draw=slLine, line width=0.4pt, inner sep=2.1pt, align=left, sh,
               font=\fontsize{6}{7}\selectfont},
  av/.style={circle, fill=#1, text=white, inner sep=0pt, minimum size=9pt, font=\fontsize{5.2}{5.2}\selectfont},
  tn/.style={font=\fontsize{6}{6.4}\selectfont\bfseries, text=slMut, inner sep=0pt},
  tag/.style={rectangle, rounded corners=2.5pt, fill=#1, text=white, inner sep=1.1pt, font=\fontsize{6}{7}\selectfont\bfseries},
  pill/.style={rectangle, rounded corners=3pt, fill=#1, text=white, inner sep=1.3pt, font=\fontsize{6}{7}\selectfont\bfseries},
  chip/.style={rectangle, rounded corners=3pt, inner sep=1.2pt, font=\fontsize{6}{7}\selectfont},
  hdr/.style={font=\fontsize{6.4}{7.4}\selectfont\bfseries, text=slInk, anchor=south west, inner sep=0pt},
  panel/.style={fill=slBg, draw=slLine!70, line width=0.4pt, rounded corners=5pt, inner sep=0pt},
]
\node[hdr] at (0.02,0.24) {{\color{slMut}\faIcon{comments}}~~Conversation \textmd{\textcolor{slMut}{(illustrative)}}};
\coordinate (r0) at (0.52,0.04);
\def\turn#1#2#3#4#5#6{%
  \node[ubub, anchor=north west, #6] (q#1) at (r#2) {#4};
  \node[av=slMut!70] at ($(q#1.west)+(-0.2,0)$) {\faIcon{user}};
  \node[tn, anchor=east] at ($(q#1.west)+(-0.38,0)$) {#3};
  \node[abub, anchor=east] (a#1) at ($(q#1.east -| 7.12,0)$) {#5};
  \node[av=pBlue!75] at ($(a#1.east)+(0.2,0)$) {\faIcon{robot}};
  \coordinate (r#1) at ($(q#1.south west)+(0,-0.08)$);}
\turn{1}{0}{1}{when did apollo 11 land on the moon?}{July 20, 1969.}{}
\turn{2}{1}{2}{who walked on it?}{Neil Armstrong first,\\then Buzz Aldrin.}{}
\turn{3}{2}{3}{where did he grow up?}{Montclair, New Jersey.}{}
\node[pill=pRed, anchor=east] at ($(a3.west)+(-0.08,0)$) {\faIcon{times}~answered for Aldrin};
\turn{4}{3}{4}{no, i meant the second guy\ldots{}\\wait, no, the first one. sorry}{Wapakoneta, Ohio.}{draw=pAmber, line width=0.6pt, fill=white}
\turn{5}{4}{5}{and what did he do before nasa?}{He was a navy pilot\\and a test pilot.}{draw=pAmber, line width=0.6pt, fill=white}
\turn{6}{5}{6}{back to the mission, where did it land?}{The Sea of Tranquility.}{draw=pAmber, line width=0.6pt, fill=white}
\turn{7}{6}{7}{sorry, i meant the next mission, 12}{The Ocean of Storms.}{draw=pAmber, line width=0.6pt, fill=white}
\node[hdr] (h2) at ($(r7 -| 0.02,0)+(0,-0.42)$) {{\color{slMut}\faIcon{search}}~~What turns 4--7 ask for};
\def\need#1#2#3#4#5{%
  \node[tn, anchor=west] (n#1) at (#2) {#1};
  \node[tag=pAmber, anchor=west, minimum width=1.45cm] (nt#1) at ($(n#1.east)+(0.08,0)$) {#3};
  \node[anchor=west, inner sep=0pt] (nq#1) at ($(nt#1.east)+(0.08,0)$) {#4};
  \coordinate (#5) at ($(n#1.west)+(0,-0.3)$);}
\def\rej#1#2{\node[chip, draw=pRed, text=pRed, fill=white, anchor=east] at ($(nq#1.west -| 7.45,0)$) {\faIcon{times}~#2};}
\need{4}{$(h2.south west -| 0.2,0)+(0,-0.2)$}{self-repair}{where did \textbf{Neil Armstrong} grow up?}{s4}\rej{4}{Buzz Aldrin}
\need{5}{s4}{follow-up}{what did \textbf{Neil Armstrong} do before NASA?}{s5}\rej{5}{Buzz Aldrin}
\need{6}{s5}{return}{where did \textbf{Apollo 11} land?}{s6}
\need{7}{s6}{value revision}{where did \textbf{Apollo 12} land?}{s7}\rej{7}{Apollo 11}
\begin{scope}[on background layer]
\coordinate (a1) at (-0.04,0.44); \coordinate (a2) at ($(r7 -| 7.62,0)+(0,0.0)$);
\coordinate (c1) at ($(h2.north west)+(-0.06,0.1)$); \coordinate (c2) at ($(s7 -| 7.62,0)+(0,0.08)$);
\node[panel, fit=(a1)(a2)] {};
\node[panel, fit=(c1)(c2)] {};
\end{scope}
\end{tikzpicture}
\caption{A running example, written for illustration. The system answers turn 3 for the wrong astronaut, and in turn 4 the user corrects it after first misspeaking. Turns 5--7 are a follow-up, a return and a value revision. To search well, a system must resolve every turn to the topic the user means (bold) and set aside the topics the user rejected ($\times$).}
\label{fig:example}
\end{figure}

Figure~\ref{fig:example} illustrates the challenge. The conversation begins with Apollo 11 and the two astronauts who walked on the Moon. When the user asks, ``where did he grow up?'', the system answers for Buzz Aldrin. The user replies, ``no, i meant the second guy\ldots{} wait, no, the first one. sorry''. This turn asks no explicit question, names no person, and immediately revises its own correction, yet it changes how the previous turn should be interpreted. The system must recover the open question, reject the earlier interpretation, and resolve ``the first one'' to Neil Armstrong. Later turns continue to shift the relevant context, from a follow-up about Armstrong to a return to the mission and then a switch to Apollo 12.

Existing methods do not explicitly represent such changes. Rewriters \citep{t5qr,convgqr,chiq,convsearchr1} are designed to resolve missing context in the current query, but here the user provides only a correction and the intended query must be recovered from an earlier turn. Conversational dense retrievers \citep{convdr,chatretriever} encode the dialogue without marking which earlier context is no longer valid. Memory-based methods such as CMT-RAG \citep{cmtrag} preserve conversational history, but do not distinguish a rejected topic from one that remains relevant.

We therefore propose \system{}, a conversational memory that later turns can revise. It reads each turn once and records its topic, associated facts, and operations such as adding, abandoning, or reactivating a topic. The status of each topic follows from these operations, allowing a dropped topic to retain its information and later become active again. A 3B writer, distilled from teacher-generated memories, updates the memory with one call per turn and produces the search query, which is combined with evidence from the memory during retrieval. In Figure~\ref{fig:method}, the writer drops Aldrin after the correction, makes Armstrong the current topic, and recovers the query ``where did Neil Armstrong grow up?''.

Evaluating this behaviour requires data that existing benchmarks do not isolate. A screen for phrases such as ``I meant'' and ``not that one'' finds none among the 47{,}964 user turns in TopiOCQA's training and validation conversations or the 6{,}480 user turns in CORAL's evaluation conversations. Although this screen captures only explicit corrections, it suggests that such turns are poorly represented in these datasets. We therefore introduce \dataset{}, a benchmark of 4{,}429 test questions over TopiOCQA, CORAL, and OR-QuAC. Its six tiers cover returns and revisions, ranging from naturally occurring returns in TopiOCQA to corrections written with language models.

We compare \system{} with seven methods: trained rewriters, a conversational dense retriever, and methods prompted on a 32B model. With a frozen retriever, \system{} reaches a mean MRR@10 of 0.701. It significantly outperforms the strongest competitor on TopiOCQA (0.686 vs.\ 0.620) and OR-QuAC (0.783 vs.\ 0.594), and achieves the highest score on CORAL (0.634), although the difference is not statistically significant. A writer distilled in the same way, but used without the memory and its search steps, trails \system{} by 0.084.

Our contributions are as follows.
\textbf{(1)}~We study revisions and returns in conversational search and introduce \dataset{}, a benchmark of 4{,}429 test questions in six tiers over three corpora (\S\ref{sec:benchmark}).
\textbf{(2)}~We propose \system{}, a revisable memory written turn by turn by a 3B writer distilled from teacher-generated memories, together with a retrieval procedure that combines the writer's query with evidence from memory (\S\ref{sec:method}).
\textbf{(3)}~We compare \system{} with seven methods, four of them also trained on our data (Appendix~\ref{app:competitors}). \system{} significantly outperforms the strongest competitor in 6 of the 12 per-tier comparisons and is statistically tied in the remainder (\S\ref{sec:results}).

\section{Related Work}
\label{sec:related}

\paragraph{Conversational query rewriting and dense retrieval.}
Conversational query rewriting turns the current question into a standalone query, with a rewriter trained on human rewrites or on retrieval feedback \citep{t5qr,convgqr,convsearchr1} or with a prompted large language model \citep{chiq}. Conversational dense retrieval instead encodes the current question and the earlier turns into a single query representation \citep{convdr,chatretriever}. Both designs read the full history at every turn and keep no state between turns. Whether an earlier request still holds must therefore be inferred again from the raw text each time, and a withdrawn request is never marked.

\paragraph{Conversational memory.}
CMT-RAG \citep{cmtrag} is the closest work to ours. At each turn it decomposes the question into sub-questions and stores each as a trace with its retrieved passages and a reader's answer; later turns retrieve these traces. The record is append-only, and each trace depends on a generated answer. Long-term memories for chat agents store facts and update them as new information arrives \citep{mem0,zep}, and Context-Agent \citep{contextagent} organises the dialogue history into topic branches. Like CMT-RAG, they provide context for a generator. Ours serves conversational search and is evaluated on retrieval. It records the conversation's topics and what each turn did to them, and writes the search query for the topic the user currently means, even after a correction or a return. Of the conversational search methods compared in Table~\ref{tab:comparison} (Appendix~\ref{app:related}), only \system{} reads each turn once, keeps a memory that later turns can revise and needs no generated answer.

\paragraph{Benchmarks.}
TopiOCQA \citep{topiocqa} contains topic switches and returns, CORAL \citep{coral} samples topic shifts through Wikipedia sections, and OR-QuAC \citep{orquac} keeps each conversation on one article. None of these papers, nor RECOR \citep{recor}, describes turns in which the user corrects or revises an earlier request.

\paragraph{Our contribution.}
Our work addresses the non-additive nature of conversational search. On the data side, we introduce \dataset{} to benchmark the user corrections, revisions and returns that existing datasets do not isolate. Methodologically, \system{} departs from full-history and append-only models by processing each turn exactly once to maintain a revisable memory graph, actively isolating dropped context as user intents evolve.
\section{The \dataset{} Benchmark}
\label{sec:benchmark}

\dataset{} evaluates whether conversational retrieval systems can maintain the correct retrieval target as user intent evolves. Each question appears as the final turn of a conversation and tests whether the system can resolve a return to an earlier topic, revise an existing interpretation, or continue from previously revised context.

\begin{table}[t]
\centering
\small
\resizebox{\columnwidth}{!}{%
\begin{tabular}{@{}lp{5.2cm}@{}}
\toprule
\textbf{Tier} & \textbf{Final turn (example)} \\
\midrule
\rowcolor{gray!12}\multicolumn{2}{c}{\textit{Returns}} \\
Return-1 & \textit{when did the talk show begin?} (\emph{The Panel}, after six turns about John Bishop) \\
Return-2+ & \textit{Sorry for the detour --- how did the one I asked about first evolve?} (the red fox, after polo and ancient Macedonia) \\
\midrule
\rowcolor{gray!12}\multicolumn{2}{c}{\textit{Revisions}} \\
Correction & \textit{no, the one arthur b. rubinstein composed the music for. not the animated film about simba} (\emph{WarGames}, after an answer about \emph{The Lion King}) \\
Follow-up & \textit{can you name more muppet characters?} (two turns after the user leaves a detour and returns to \emph{Sesame Street}) \\
Self-repair & \textit{no, i meant the whatever it takes guy\ldots{} sorry, no, the actor who named physical comedy as his big inspiration} (Bryan Cranston, not Dick Van Dyke) \\
Value revision & \textit{hmm, i meant the one in 1863} (from \emph{War of 1812} to \emph{War of 1863}) \\
\bottomrule
\end{tabular}}
\caption{The six tiers of \dataset{}, each with a final turn from the test set. For each example, \system{} ranks a gold passage in the top two for all three seeds, while the five strongest competitors miss the top 10.}
\label{tab:tiers}
\end{table}

We build \dataset{} from TopiOCQA \citep{topiocqa}, CORAL \citep{coral}, and OR-QuAC \citep{orquac}. Each conversational topic corresponds to a Wikipedia page. For TopiOCQA and OR-QuAC, the retrieval collection contains passages from that page, while CORAL uses passages from web pages cited by the corresponding Wikipedia article. For each question, \textbf{gold passages} answer the question for the intended topic, while for written questions \textbf{stale passages} belong to context that is no longer relevant, such as a rejected topic, a topic the user did not return to, or the original page before a value revision.

\subsection*{Tiers}
\dataset{} defines six tiers (Table~\ref{tab:tiers}) that capture two forms of conversational change: returning to earlier context and revising the current context.
\textbf{Return-1} resumes the topic discussed immediately before the current one, while \textbf{Return-2+} returns to a topic at least two topics earlier. In both cases, the target is referred to indirectly through a pronoun or description rather than named explicitly, requiring the system to recover it from the dialogue history.

The remaining tiers capture different forms of revision. \textbf{Correction} rejects one or more topics and redirects the conversation to the intended one. \textbf{Follow-up} is a short question whose interpretation depends on a preceding correction or corrective return. \textbf{Self-repair} occurs when the user introduces a correction and then revises it within the same turn \citep{schegloff1977}. \textbf{Value revision} changes a topic-defining value, shifting the retrieval target to a closely related page with a nearly identical name.

\subsection*{Construction}
\dataset{} builds on conversations from TopiOCQA, CORAL and OR-QuAC. The natural returns are original TopiOCQA turns; all other revision turns were generated with GPT-5.4-mini, refined with Claude Fable 5.1 and then verified individually. We construct \dataset{} in four main ways.
\begin{enumerate}[nosep,leftmargin=*]
\item \textbf{Natural returns} use TopiOCQA turns that resume a topic discussed earlier in the same conversation.
\item \textbf{Built revisions} start from two topics of the same type within a TopiOCQA conversation. We add a question that applies to both, initially answer it for the wrong topic, and then redirect the conversation through a correction, self-repair, or correction followed by a short follow-up. Each example is paired with another that shares the same history but targets the alternative topic, preventing topic position from revealing the answer.
\item \textbf{Spliced conversations} take the opening of a source conversation, insert two intervening topics, and end with a return or correction. This creates histories in which the intended topic must be recovered across competing context.
\item \textbf{Value revisions} extend a conversation to a closely related page whose title differs in a value, such as a number, year, or edition. The final turn changes that value and redirects the retrieval target to the corresponding page.
\end{enumerate}

\subsection*{Size and Splits}
The test set contains \textbf{4{,}429 questions}: 3{,}500 from 1{,}196 TopiOCQA conversations, 617 from 268 CORAL conversations, and 312 from 219 OR-QuAC conversations. TopiOCQA covers all six tiers, while CORAL and OR-QuAC cover corrections and returns. Of the TopiOCQA questions, 2{,}839 (81\%) originate from conversations in its official training split, so models trained on TopiOCQA may have prior exposure to these dialogue histories.

The training set contains \textbf{1{,}790 questions from 769 conversations}, including 839 from TopiOCQA and 951 from CORAL. No conversation overlaps with the test set, and OR-QuAC is not used for training.

\section{\system{}}
\label{sec:method}

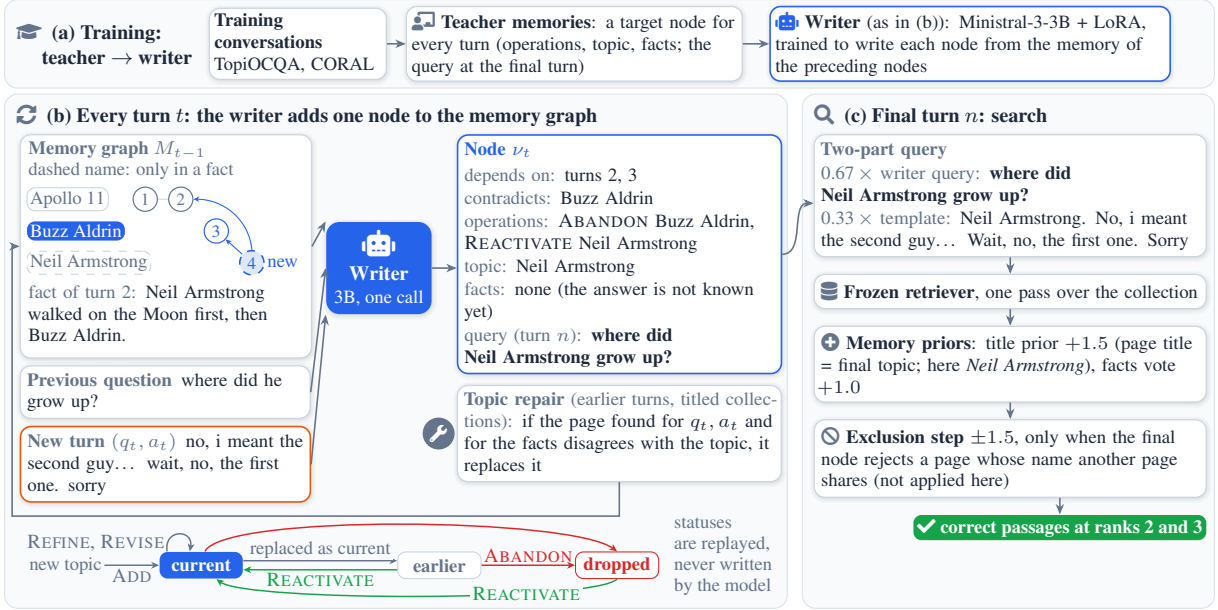
\begin{figure*}[t]
\centering
\begin{tikzpicture}[
  font=\fontsize{7}{8.2}\selectfont, text=slInk, >={Stealth[length=4pt,width=3.2pt]},
  sh/.style={blur shadow={shadow blur steps=6, shadow xshift=0.25pt, shadow yshift=-0.5pt, shadow blur extra rounding=1pt, shadow opacity=18}},
  card/.style={rectangle, rounded corners=4pt, fill=white, draw=slLine, line width=0.45pt, inner sep=2.6pt, align=left, sh, anchor=north west},
  ico/.style={circle, fill=#1, text=white, inner sep=0pt, minimum size=12pt, font=\fontsize{6.5}{6.5}\selectfont},
  pill/.style={rectangle, rounded corners=3pt, fill=#1, text=white, inner sep=1.4pt, font=\fontsize{7}{8}\selectfont\bfseries},
  chip/.style={rectangle, rounded corners=3pt, inner sep=1.2pt, font=\fontsize{7}{8}\selectfont},
  hdr/.style={font=\fontsize{7.6}{8.6}\selectfont\bfseries, text=slInk, anchor=north west, inner sep=0pt},
  flow/.style={->, draw=slMut, line width=0.6pt},
  opf/.style={font=\fontsize{7.5}{8.2}\selectfont},
]
\def\op#1{{\fontsize{7.5}{8.2}\selectfont\textsc{#1}}}
\fill[slBg, draw=slLine!70, line width=0.4pt, rounded corners=5pt] (0,-2.55) rectangle (10.35,4.25);
\fill[slBg, draw=slLine!70, line width=0.4pt, rounded corners=5pt] (10.55,-2.55) rectangle (16.0,4.25);
\node[hdr] at (0.15,4.12) {{\color{slMut}\faIcon{sync-alt}}~~(b)~Every turn $t$: the writer adds one node to the memory graph};
\node[hdr] at (10.7,4.12) {{\color{slMut}\faIcon{search}}~~(c)~Final turn $n$: search};
\fill[slBg, draw=slLine!70, line width=0.4pt, rounded corners=5pt] (0,4.33) rectangle (16.0,5.5);
\begin{scope}[tc/.style={card, anchor=west, font=\fontsize{7}{8.2}\selectfont, inner sep=2pt}]
\node[hdr, anchor=west, align=left] at (0.15,4.915) {{\color{slMut}\faIcon{graduation-cap}}~~(a)~Training:\\[1pt]\hspace*{1.25em}teacher $\rightarrow$ writer};
\node[tc, text width=2.2cm] (tdat) at (2.7,4.915) {\textbf{Training}\\\textbf{conversations}\\TopiOCQA, CORAL};
\node[tc, text width=4.3cm] (tteach) at ($(tdat.east)+(0.25,0)$) {{\color{slMut}\faIcon{chalkboard-teacher}}~\textbf{Teacher memories}: a target node for every turn (operations, topic, facts; the query at the final turn)};
\node[tc, text width=5.15cm, draw=pBlue, line width=0.6pt] (tstud) at ($(tteach.east)+(0.35,0)$) {{\color{pBlue}\faIcon{robot}}~\textbf{Writer} (as in (b)): \mbox{Ministral-3-3B} + LoRA, trained to write each node from the memory of the preceding nodes};
\draw[flow] (tdat.east) -- (tteach.west);
\draw[flow] (tteach.east) -- (tstud.west);
\end{scope}
\node[card, minimum width=3.85cm, minimum height=2.95cm] (mem) at (0.2,3.72) {};
\begin{scope}[tn/.style={circle, draw=#1, fill=white, line width=0.45pt, inner sep=0pt, minimum size=0.32cm, font=\fontsize{6.5}{6.5}\selectfont, text=#1},
  lane/.style={draw=slLine, line width=0.35pt}, lab/.style={chip, font=\fontsize{7}{7.6}\selectfont, align=left, anchor=west},
  be/.style={->, draw=pBlue, line width=0.4pt, >={Stealth[length=2.2pt,width=2pt]}}]
\node[anchor=north west, inner sep=0pt, align=left] at (0.31,3.63) {\textcolor{slMut}{\textbf{Memory graph} $M_{t-1}$}\\[-0.5pt]\textcolor{slMut}{dashed name: only in a fact}};
\def\cx#1{\fpeval{1.86+(#1-1)*0.47}}
\node[lab, draw=slLine, text=slMut, fill=white] at (0.29,2.86) {Apollo 11};
\node[lab, fill=pBlue, text=white] at (0.29,2.44) {Buzz Aldrin};
\node[lab, draw=slLine, text=slMut, fill=white, dashed] at (0.29,2.02) {Neil Armstrong};
\draw[lane] (\cx{1},2.86) -- (\cx{2},2.86);
\foreach \k in {1,2} \node[tn=slMut] (n\k) at (\cx{\k},2.86) {\k};
\node[tn=pBlue] (n3) at (\cx{3},2.44) {3};
\node[tn=pBlue, dashed, fill=pBlue!15, line width=0.6pt] (n4) at (\cx{4},2.02) {4};
\node[font=\fontsize{7}{8}\selectfont, text=pBlue, anchor=west, inner sep=0.5pt] at ($(n4.east)+(0.01,0)$) {new};
\draw[be] (n4) -- (n3);
\draw[be] (n4.north) to[out=95, in=0] (n2.east);
\node[anchor=north west, text width=3.6cm, inner sep=0pt] at (0.31,1.73)
  {\textcolor{slMut}{fact of turn 2:} \mbox{Neil Armstrong} walked on the Moon first, then \mbox{Buzz Aldrin}.};
\end{scope}
\node[card, text width=3.65cm] (prev) at ($(mem.south west)+(0,-0.09)$) {%
  \textcolor{slMut}{\textbf{Previous question}}\enspace where did he grow up?};
\node[card, text width=3.65cm, draw=pAmber, line width=0.6pt] (new) at ($(prev.south west)+(0,-0.09)$) {%
  \textcolor{slMut}{\textbf{New turn} $(q_t, a_t)$}\enspace no, i meant the second guy\ldots{} wait, no, the first one. sorry};
\node[rectangle, rounded corners=5pt, fill=pBlue, text=white, align=center, inner sep=3pt, minimum width=1.3cm, sh] (wr) at (4.95,1.95)
  {\fontsize{10}{10}\selectfont\faIcon{robot}\\[2pt]{\fontsize{7.6}{8.6}\selectfont\textbf{Writer}}\\[0.5pt]{\fontsize{7}{8}\selectfont 3B, one call}};
\draw[flow] (mem.east) -- (wr.north west);
\draw[flow] (prev.east) -- (wr.west);
\draw[flow] (new.east) -- (wr.south west);
\node[card, text width=4.1cm, draw=pBlue, line width=0.6pt] (node) at (5.98,3.72) {%
  \textcolor{pBlue}{\textbf{Node} $\nu_t$}\\[1.5pt]
  \textcolor{slMut}{depends on:} turns 2, 3\\[0.5pt]
  \textcolor{slMut}{contradicts:} \mbox{Buzz Aldrin}\\[0.5pt]
  \textcolor{slMut}{operations:} \op{Abandon} Buzz Aldrin, \mbox{\op{Reactivate}} Neil Armstrong\\[0.5pt]
  \textcolor{slMut}{topic:} Neil Armstrong\\[0.5pt]
  \textcolor{slMut}{facts:} none (the answer is not known yet)\\[1pt]
  \textcolor{slMut}{query (turn $n$):} \textbf{where did \mbox{Neil Armstrong} grow up?}};
\draw[flow] (wr.east) -- (wr.east -| node.west);
\node[card, text width=4.1cm] (rep) at ($(node.south west)+(0,-0.14)$) {%
  \textcolor{slMut}{\textbf{Topic repair}} \textcolor{slMut}{(earlier turns, titled collections):} if the page found for $q_t, a_t$ and for the facts disagrees with the topic, it replaces it};
\node[ico=slMut] at ($(rep.west)+(-0.24,0)$) {\faIcon{wrench}};
\coordinate (lb) at (rep.south |- 0,-1.33);
\draw[flow] (rep.south) -- (lb) -- (0.1,-1.33) |- (mem.west);
\begin{scope}[st/.style={chip, font=\fontsize{7}{8}\selectfont\bfseries, minimum height=10pt, minimum width=1.1cm, align=center},
  op/.style={font=\fontsize{7.5}{8.2}\selectfont, inner sep=0.8pt},
  tr/.style={->, line width=0.55pt, >={Stealth[length=3pt,width=2.6pt]}}]
\node[st, fill=pBlue, text=white] (cur) at (2.6,-1.98) {current};
\node[st, draw=slLine, fill=white, text=slMut] (ear) at (5.75,-1.98) {earlier};
\node[st, draw=pRed, fill=white, text=pRed] (drp) at (8.1,-1.98) {dropped};
\node[op, text=slMut, anchor=east, font=\fontsize{7}{8}\selectfont] (nt) at (1.32,-1.98) {new topic};
\draw[tr, slMut] (nt.east) -- node[op, below, text=slMut] {\textsc{Add}} (cur.west);
\draw[tr, slMut] ($(cur.north)+(-0.45,0)$) .. controls +(-0.05,0.3) and +(0.05,0.3) .. ($(cur.north)+(-0.15,0)$)
  node[op, pos=0.15, left, text=slMut] {\textsc{Refine}, \textsc{Revise}};
\draw[tr, pRed] ($(cur.north)+(0.05,0)$) .. controls +(0.25,0.5) and +(-0.6,0.45) .. (drp.north);
\draw[tr, slMut] ($(cur.east)+(0,0.06)$) -- node[op, above, text=slMut, fill=slBg, font=\fontsize{7}{8}\selectfont] {replaced as current} ($(ear.west)+(0,0.06)$);
\draw[tr, pGreen] ($(ear.west)+(0,-0.06)$) -- node[op, below, text=pGreen] {\textsc{Reactivate}} ($(cur.east)+(0,-0.06)$);
\draw[tr, pRed] (ear.east) -- node[op, above, text=pRed] {\textsc{Abandon}} (drp.west);
\draw[tr, pGreen] (drp.south) .. controls +(-0.3,-0.3) and +(0.4,-0.3) .. node[op, pos=0.3, fill=slBg, text=pGreen] {\textsc{Reactivate}} ($(cur.south)+(0.2,0)$);
\node[anchor=west, text=slMut, align=left, font=\fontsize{7}{8}\selectfont] at (8.72,-1.86) {statuses\\ are replayed,\\ never written\\ by the model};
\end{scope}
\node[card, text width=5.0cm] (q) at (10.7,3.72) {%
  \textcolor{slMut}{\textbf{Two-part query}}\\[1pt]
  \textcolor{slMut}{0.67\,$\times$ writer query:} \textbf{where did \mbox{Neil Armstrong} grow up?}\\[0.5pt]
  \textcolor{slMut}{0.33\,$\times$ template:} Neil Armstrong. No, i meant the second guy\ldots{} Wait, no, the first one. Sorry};
\draw[flow] (node.east) to[out=0, in=180] ($(q.west)+(0,-0.1)$);
\node[card, text width=5.0cm] (ret) at ($(q.south west)+(0,-0.22)$) {%
  {\color{slMut}\faIcon{database}}~\textbf{Frozen retriever}, one pass over the collection};
\node[card, text width=5.0cm] (pri) at ($(ret.south west)+(0,-0.22)$) {%
  {\color{slMut}\faIcon{plus-circle}}~\textbf{Memory priors}: title prior $+1.5$ (page title = final topic; here \emph{Neil Armstrong}), facts vote $+1.0$};
\node[card, text width=5.0cm] (exc) at ($(pri.south west)+(0,-0.22)$) {%
  {\color{slMut}\faIcon{ban}}~\textbf{Exclusion step} $\pm1.5$, only when the final node rejects a page whose name another page shares (not applied here)};
\node[pill=pGreen, anchor=north east] (rk) at ($(exc.south east)+(0,-0.22)$) {\faIcon{check}~correct passages at ranks 2 and 3};
\foreach \a/\b in {q/ret, ret/pri, pri/exc} \draw[flow] (\a.south) -- (\b.north);
\draw[flow] (exc.south -| rk.north) -- (rk.north);
\end{tikzpicture}
\caption{\system{}. (a)~The 3B writer is distilled from teacher-written target memories. (b)~At every turn it adds one node to the memory graph; statuses follow from the operations. (c)~At the final turn the node's query is scored with a template query and evidence from the memory. Example: turn 4 of the conversation in Figure~\ref{fig:example}, searched as the final turn. The memory of turns 1--3 is shown as intended, with turn 3 stored as the system's answer about Buzz Aldrin; the node of turn 4 is the writer's output for this memory (seed 3).}
\label{fig:method}
\end{figure*}

\system{} models a conversation as a revisable memory that is updated once per turn. Instead of repeatedly encoding the full dialogue history, it preserves the information needed for future turns in a structured memory and revises that state as the conversation develops (Figure~\ref{fig:method}).

\subsection{Task Formulation}
\label{sec:task}

A conversation consists of turns $(q_t, a_t)$ for $t = 1, \ldots, n-1$, followed by a final question $q_n$ whose answer is unknown. Given a passage collection $\mathcal{P}$, the task is to rank passages for $q_n$.

Conventional query rewriting derives a standalone query from the full dialogue history. \system{} instead maintains a memory $M_t$ and updates it once per turn:
\begin{equation}
M_t = \mathrm{update}\big(M_{t-1},\, q_{t-1},\, (q_t, a_t)\big),
\label{eq:update}
\end{equation}
where $a_t$ is empty at the final turn. The final update also produces the search query. Earlier turns are therefore carried forward through memory rather than reread from the raw conversation.

\subsection{A Revisable Memory}
\label{sec:memory}

After turn $t$, the memory contains one node per turn, $M_t = \{\nu_1, \ldots, \nu_t\}$, where $\nu_t = (O_t, e_t, F_t)$ stores the turn's operations $O_t$, topic $e_t$, and facts $F_t$. Each node also records the earlier turns it depends on and the topics it contradicts, forming a graph over the conversation.

\system{} uses five operations to represent how a turn changes the conversational state: \textsc{Add} introduces a new topic, \textsc{Refine} continues the current one, \textsc{Revise} changes an established value, \textsc{Abandon} drops a topic, and \textsc{Reactivate} restores an earlier topic.\footnote{The writer's prompt (Figure~\ref{fig:prompt}) also defines \textsc{Retract}, for a premise declared false; no training target contains it, and the writer never produces it.}

Each topic has one of three states: \textsc{current}, \textsc{earlier}, or \textsc{dropped}. These states are not predicted directly; they are derived by replaying the operations in turn order. This keeps dropped topics and their facts in memory, so later turns can restore them without reconstructing the earlier dialogue.

\subsection{Writing the Memory}
\label{sec:student}

A small language model, the memory writer, performs each update in a single call. At turn $t$, it receives the current memory as text, the previous question, and the new turn. It predicts the dependencies, contradictions, and operations before assigning the topic and storing any new facts.

At the final turn, the writer also produces the search query. It preserves the user's information need while resolving pronouns, descriptions, and references through the memory. If the turn only corrects an earlier interpretation, the writer carries forward the unresolved question and applies it to the intended topic. References to entities mentioned in stored facts are resolved in the same way.

The writer is trained through distillation from teacher-generated memories. Target nodes are built from GPT-5.4-mini annotations corrected with Claude Opus 5.5, and a Ministral-3-3B \citep{ministral} model with a LoRA adapter \citep{lora} is trained to reproduce them. For part of the training data, the input memory comes from an auxiliary writer trained on different conversations, exposing the model to imperfect memory states similar to those encountered at inference time (Appendix~\ref{app:system}).

\subsection{Grounding Topics in the Corpus}
\label{sec:repair}

The writer can occasionally attach a turn to the wrong topic, especially when the conversation shifts to an entity introduced in the previous answer. For collections with page titles, \system{} uses topic repair to ground the predicted topic against the corpus.

After each history turn, it retrieves with the turn's question and answer and separately with the node's stored facts. If both searches identify the same page and that page conflicts with the writer's topic, the node is reassigned to the retrieved page. CORAL has no page titles \citep{coral}, so its memories retain the writer's original topic assignments.

\subsection{Search}
\label{sec:search}

At the final turn, \system{} searches with the frozen Reason-ModernColBERT retriever \citep{colbert,reasonmoderncolbert}. Retrieval combines two complementary queries: the writer's resolved query $q^{s}$ and a template query $q^{t}$ that retains the user's wording while foregrounding the current topic. For passage $p$,
\begin{equation}
s(p) = z\big(0.67\, S(q^{s}, p) + 0.33\, S(q^{t}, p)\big),
\label{eq:score}
\end{equation}
where $S(q, p)$ is the late-interaction score, normalised by the query's total token weight, and $z(\cdot)$ standardises scores over the collection. The writer query expresses the resolved information need, while the template provides a complementary signal when the generated query drifts.

When page titles are available, \system{} also uses memory-derived ranking signals. A title prior increases the score of passages whose page title matches the final topic, while a facts vote favours pages supported by the facts stored for that topic. Final turns that only replace a number are handled by a small rule that updates the number in the topic and query before retrieval.

Finally, revisions may distinguish between different pages with the same surface name. In such cases, \system{} uses the memory to identify the rejected page and adjusts the ranking against it while favouring same-name alternatives. This exclusion step follows EXCISE \citep{excise} and leaves the underlying retriever frozen.

\section{Experimental Setup}
\label{sec:setup}

\subsection*{Competitors and Retrieval Architecture}
We compare \system{} against seven competitors representing diverse conversational search approaches: trained query rewriters (T5QR, ConvGQR$^{\dagger}$, ConvSearch-R1), prompted query rewriters (LLM-Rewrite, CHIQ-AD), trace-based multi-turn memory (CMT-RAG$^{\ddagger}$), and conversational dense retrieval (ChatRetriever).

All query-based methods, including \system{}, use the same frozen Reason-ModernColBERT retriever over identical collections. \system{} additionally incorporates the memory-guided search steps detailed in \S\ref{sec:search}, whereas ChatRetriever relies on its own conversational encoder. An \textit{Original Query} baseline is included to provide a reference without context by searching only with the final turn.

\subsection*{Training and Implementation Details}
To provide a comprehensive comparison, we evaluate these competitors in their off-the-shelf configurations and, where applicable, fine-tune or train them on \dataset{}'s training questions (T5QR-FT, ConvSearch-R1-FT, ChatRetriever-FT, and ConvGQR$^{\dagger}$). All prompted competitors---LLM-Rewrite, CHIQ-AD, and CMT-RAG$^{\ddagger}$---run on Qwen2.5-32B-Instruct \citep{qwen25}. Methods marked with $^{\dagger}$ or $^{\ddagger}$ denote reimplementations; complete configuration details, including specific hyperparameter settings and deviations from the original systems, are provided in Appendix~\ref{app:competitors}.

\subsection*{Evaluation Protocol}
The primary metric is MRR@10, supplemented by NDCG@3 and Recall@10/100 (Appendix~\ref{app:results}). Overall performance is the unweighted mean across the three test sets, and tier-level comparisons are restricted to subsets with at least 30 questions. For revision questions, we also track whether a stale passage from a rejected request outranks all gold passages.

Statistical significance is assessed via paired bootstrap resampling over conversations. We report a 95\% interval for the mean per-question difference between \system{} and each competitor, and treat the two as statistically tied when the interval contains zero. All design choices for \system{} and the competitors were made on six development sets, and no model, including \system{}, is trained or tuned on the test set. \system{} results reflect the mean of three random seeds, while competitors are run once, with CHIQ-AD and ConvSearch-R1 sampling at a temperature of 0.7, as in their respective papers.

\section{Results and Analysis}
\label{sec:results}

\begin{table}[t]
\centering
\setlength{\tabcolsep}{3pt}
\resizebox{\columnwidth}{!}{%
\begin{tabular}{@{}llcccc@{}}
\toprule
\textbf{Method} & \textbf{Size} & \textbf{TopiOCQA} & \textbf{CORAL} & \textbf{OR-QuAC} & \textbf{Mean} \\
\midrule
Original Query & -- & 0.074 & 0.360 & 0.263 & 0.232 \\
\midrule
\rowcolor{gray!12}\multicolumn{6}{c}{\textit{Trained rewriters}} \\
T5QR & 0.2B & 0.243 & 0.460 & 0.353 & 0.352 \\
T5QR-FT & 0.2B & 0.342 & 0.469 & 0.302 & 0.371 \\
ConvGQR$^{\dagger}$ & 0.2B & 0.288 & 0.404 & 0.470 & 0.387 \\
ConvSearch-R1 & 3B & 0.591 & 0.546 & 0.572 & 0.570 \\
ConvSearch-R1-FT & 3B & 0.611 & 0.570 & 0.588 & \underline{0.589} \\
\midrule
\rowcolor{gray!12}\multicolumn{6}{c}{\textit{Prompted methods}} \\
LLM-Rewrite & 32B & 0.499 & 0.609\rlap{$^{\circ}$} & 0.581 & 0.563 \\
CHIQ-AD & 32B & 0.436 & \underline{0.617}\rlap{$^{\circ}$} & \underline{0.594} & 0.549 \\
CMT-RAG$^{\ddagger}$ & 32B & 0.283 & 0.507 & 0.440 & 0.410 \\
\midrule
\rowcolor{gray!12}\multicolumn{6}{c}{\textit{Conversational dense retrieval}} \\
ChatRetriever & 7B & 0.448 & 0.433 & 0.468 & 0.450 \\
ChatRetriever-FT & 7B & \underline{0.620} & 0.524 & 0.548 & 0.564 \\
\midrule
\textbf{\system{}} (ours) & 3B & \textbf{0.686} & \textbf{0.634} & \textbf{0.783} & \textbf{0.701} \\
\bottomrule
\end{tabular}}
\caption{MRR@10 on the three test sets with the frozen retriever. \system{} is the mean of three seeds, which differ by at most 0.009 on any set. Best in \textbf{bold}, second best \underline{underlined}. $^{\circ}$: the 95\% interval of \system{} minus this method contains zero (paired bootstrap over conversations); every other difference is significant. ChatRetriever rows use their own retriever. -FT: continued on our training data; $^{\dagger}$: trained on our data, no released model; $^{\ddagger}$: our re-implementation.}
\label{tab:main_results}
\end{table}

\subsection*{Overall Retrieval Effectiveness}
\system{} achieves the highest MRR@10 across all three evaluation corpora (Table~\ref{tab:main_results}). It achieves an overall mean MRR@10 of 0.701, which exceeds the strongest competitor, ConvSearch-R1-FT (0.589). At the dataset level, \system{} scores 0.686 on TopiOCQA, 0.634 on CORAL, and 0.783 on OR-QuAC. It establishes significant margins over the top-performing competitors on TopiOCQA and OR-QuAC, beating ChatRetriever-FT by $+0.066$ and CHIQ-AD by $+0.190$, respectively. On CORAL, \system{} statistically ties with the leading prompted methods. Crucially, \system{} maintains this superiority on OR-QuAC despite the corpus being entirely excluded from its training data, which indicates robust generalisation to a corpus not seen in training. \system{} is also best on NDCG@3 and keeps the best mean MRR@10 with two other retrievers (Appendix~\ref{app:results}).

\begin{table}[t]
\centering
\setlength{\tabcolsep}{3pt}
\resizebox{\columnwidth}{!}{%
\begin{tabular}{@{}lrcclr@{}}
\toprule
\textbf{Tier} & \textbf{$n$} & \textbf{\system{}} & \textbf{Best} & \textbf{Competitor} & \textbf{$\Delta$} \\
\midrule
\rowcolor{gray!12}\multicolumn{6}{c}{\textit{TopiOCQA}} \\
Correction     & 759  & 0.796 & 0.750 & CR-FT   & $+0.046^{*}$ \\
Self-repair    & 608  & 0.796 & 0.772 & CR-FT   & $+0.024$ \\
Value revision & 233  & 0.645 & 0.504 & CR-FT   & $+0.141^{*}$ \\
Follow-up      & 300  & 0.756 & 0.718 & CS-FT   & $+0.038$ \\
Return-1       & 1{,}028 & 0.578 & 0.576 & CS   & $+0.002$ \\
Return-2+      & 572  & 0.599 & 0.557 & CS-FT   & $+0.042^{*}$ \\
\midrule
\rowcolor{gray!12}\multicolumn{6}{c}{\textit{CORAL}} \\
Correction     & 262  & 0.681 & 0.639 & CHIQ-AD & $+0.042^{*}$ \\
Return-1       & 167  & 0.743 & 0.732 & T5QR-FT & $+0.011$ \\
Return-2+      & 188  & 0.471 & 0.504 & CHIQ-AD & $-0.033$ \\
\midrule
\rowcolor{gray!12}\multicolumn{6}{c}{\textit{OR-QuAC}} \\
Correction     & 218  & 0.785 & 0.587 & CS-FT   & $+0.198^{*}$ \\
Return-1       & 38   & 0.836 & 0.588 & CR-FT   & $+0.249^{*}$ \\
Return-2+      & 55   & 0.752 & 0.780 & CS      & $-0.029$ \\
\bottomrule
\end{tabular}}
\caption{Per-tier MRR@10 on the test sets against the strongest competitor of each tier. $\Delta$: \system{} minus that competitor; $^{*}$: the 95\% interval excludes zero (paired bootstrap over conversations). CR: ChatRetriever; CS: ConvSearch-R1.}
\label{tab:tiers_results}
\end{table}

\subsection*{Navigating Conversational Topic Shifts}
The retrieval strength of \system{} is driven by its writer, which is a distilled 3B language model that explicitly tracks the conversational state by updating a memory graph at each turn. To rigorously evaluate how well the system adapts to topic shifts and revisions, Table~\ref{tab:tiers_results} compares \system{} against whichever competitor achieved the highest score for each specific tier. Even under this highly competitive setup, \system{} significantly wins 6 of the 12 tier-level comparisons and ties the remaining 6. The largest gains appear where the target of the search changes, on value revisions in TopiOCQA ($+0.141$) and on corrections and returns to the previous topic in OR-QuAC ($+0.198$ and $+0.249$).

\providecommand{\opc}[1]{\cellcolor{pBlue!\fpeval{round(#1*0.45,0)}}#1}
\providecommand{\opf}[1]{\cellcolor{pBlue!\fpeval{round(#1*0.45,0)}}{\setlength{\fboxsep}{0.8pt}\setlength{\fboxrule}{0.7pt}\fcolorbox{navyC}{pBlue!\fpeval{round(#1*0.45,0)}}{\textbf{#1}}}}
\begin{table}[t]
\centering
\footnotesize
\setlength{\tabcolsep}{3.0pt}
\begin{tabular}{@{}lrrrrr@{}}
\toprule
\textbf{Tier} ($n$) & \textsc{Add} & \textsc{Ref.} & \textsc{Rev.} & \textsc{Aband.} & \textsc{React.} \\
\midrule
Return-1 (475)       & \opc{11.5} & \opc{10.7} & \opc{0.0}  & \opc{0.8}  & \opf{88.4} \\
Return-2+ (440)      & \opc{8.9}  & \opc{2.2}  & \opc{0.0}  & \opc{1.4}  & \opf{97.1} \\
Correction (329)     & \opc{9.1}  & \opc{14.7} & \opc{0.1}  & \opf{96.4} & \opc{85.1} \\
Self-repair (42)     & \opc{4.8}  & \opc{34.9} & \opc{0.8}  & \opf{96.8} & \opc{59.5} \\
Follow-up (69)       & \opc{5.3}  & \opf{96.1} & \opc{0.0}  & \opc{0.0}  & \opc{0.0}  \\
Value revision (128) & \opc{0.0}  & \opc{0.8}  & \opf{97.9} & \opc{99.7} & \opc{0.0}  \\
\midrule
Earlier-turn nodes   & \opc{44.9} & \opc{54.6} & \opc{0.0}  & \opc{0.5}  & \opc{2.8}  \\
\bottomrule
\end{tabular}
\caption{Operations in the writer's final node, by tier (development sets; three seeds pooled; \% of questions). A node can hold several operations. Framed: the operation that defines each tier. Last row: nodes of earlier turns.}
\label{tab:ops}
\end{table}

Because the writer accurately flags state operations, the system effectively isolates rejected context. On the development sets, the writer applies \textsc{Reactivate} on 88\% of immediate returns, \textsc{Abandon} on 96\% of corrections, and \textsc{Revise} on 98\% of value revisions (Table~\ref{tab:ops}). As a result, \system{} successfully suppresses stale information. On TopiOCQA, passages from withdrawn requests outrank every gold passage in only 5.1\% of corrections and 7.0\% of self-repairs. This represents a clear reduction in errors compared to the competing methods, whose lowest rates on these tiers are 7.0\% and 13.5\%.

\subsection*{The Impact of Memory versus Model Distillation}
To determine whether these gains stem from the explicit memory graph or simply the strength of the distilled model, we trained a memory-free writer with the same 3B model and recipe, which generates queries directly from the raw conversational text without memory or memory-guided search steps. This writer achieves a mean MRR@10 of 0.617, surpassing all competitors in Table~\ref{tab:main_results} and confirming the effectiveness of the distillation recipe. Nevertheless, integrating the explicit memory framework of \system{} yields a further significant improvement of $+0.084$ $[+0.069, +0.100]$ (Table~\ref{tab:control} in Appendix~\ref{app:results}). This margin shows that while distillation provides a strong baseline, the memory and the search steps built on it are the primary driver of the system's leading performance.

\begin{table}[t]
\centering
\setlength{\tabcolsep}{3pt}
\resizebox{\columnwidth}{!}{%
\begin{tabular}{@{}lccccc@{}}
\toprule
\textbf{Variant} & \textbf{Mean} & \textbf{All} & \textbf{Returns} & \textbf{Corr.} & \textbf{Value} \\
\midrule
Full \system{}                 & 0.662 & -- & -- & -- & -- \\
\midrule
\rowcolor{gray!12}\multicolumn{6}{c}{\textit{Query and search components}} \\
Template only & 0.591 & $-0.076^{*}$ & $-0.049^{*}$ & $-0.227^{*}$ & $-0.012$ \\
Writer query only & 0.657 & $-0.010^{*}$ & $-0.014^{*}$ & $-0.007$ & $-0.005$ \\
No priors     & 0.652 & $-0.012^{*}$ & $-0.011^{*}$ & $-0.020^{*}$ & $+0.004^{*}$ \\
No topic repair                 & 0.660 & $-0.003^{*}$ & $-0.003^{*}$ & $-0.004$ & $+0.001$ \\
No exclusion step               & 0.652 & $-0.003^{*}$ & $+0.000$ & $+0.002^{*}$ & $-0.045^{*}$ \\
\midrule
\rowcolor{gray!12}\multicolumn{6}{c}{\textit{Memory horizon: keep only the last $k$ turns}} \\
$k=4$ & 0.560 & $-0.113^{*}$ & $-0.139^{*}$ & $-0.180^{*}$ & $+0.005$ \\
$k=2$ & 0.454 & $-0.236^{*}$ & $-0.323^{*}$ & $-0.305^{*}$ & $-0.005$ \\
$k=1$ & 0.436 & $-0.256^{*}$ & $-0.349^{*}$ & $-0.331^{*}$ & $-0.015$ \\
\bottomrule
\end{tabular}}
\caption{Ablations on the six development sets (writers not retrained). Mean: six-set mean MRR@10. Pooled columns: per-question difference to the full system over all development questions, overall and for returns, corrections (incl.\ self-repair) and value revisions; $^{*}$: the 95\% interval excludes zero; --: reference row.}
\label{tab:ablation}
\end{table}

\subsection*{Query Fusion and Memory Horizons}
To identify which specific components drive the performance of the system, we conducted a targeted ablation study on the development sets (Table~\ref{tab:ablation}). During the retrieval phase, \system{} fuses two complementary signals: the explicitly generated query from the writer and a template query. The writer query expresses the resolved information need, while the template query preserves the exact wording of the user and foregrounds the current topic. When we isolate these components, the explicitly generated writer query proves to be the most critical factor. If the system relies solely on the template, correction performance drops heavily by a margin of 0.227.

Furthermore, maintaining a deep conversational history is essential for resolving returns and corrections. Truncating the memory graph to only the most recent turns results in a sharp decrease in development MRR@10. For example, limiting the history to a single turn drops the overall score from 0.662 to 0.436, and this severely degrades performance specifically on returns and corrections, while value revisions and natural CORAL questions change by at most 0.015 and 0.004 (Table~\ref{tab:ablation_groups} in Appendix~\ref{app:results}).

\subsection*{Computational Efficiency and Context Dependency}
We also investigated whether the language model needs to process the entire raw dialogue history at each step to generate accurate memory states and queries. Modifying the writer to read the full raw conversation at every turn yields no performance benefit, resulting in a score difference of $+0.000$ $[-0.007, +0.008]$. This confirms that our sequential approach, which processes each turn exactly once to update a persistent memory graph, loses no accuracy.

Additionally, we evaluated how the system handles simple queries that do not require previous context. These self-contained questions, for which the final question alone retrieves a gold passage in the top 10, make up 52\% of the CORAL test questions, against 14\% on TopiOCQA and 42\% on OR-QuAC. \system{} remains the top-performing method on these queries on every set (0.840 vs 0.767 on TopiOCQA, 0.826 vs 0.804 on CORAL, 0.894 vs 0.788 on OR-QuAC). This shows that maintaining a persistent memory graph does not degrade performance on simple and standalone searches. On the questions that need the conversation, \system{} leads by $+0.044$ on TopiOCQA and $+0.203$ on OR-QuAC and ties on CORAL, whose passages carry no page titles for the memory-based search steps to use.

\subsection*{Comparing Revisable Graphs with Trace Memory}
Compared to the closest competing paradigm of multi-turn retrieval augmented generation, \system{} demonstrates clear advantages in both accuracy and the number of model calls. The CMT-RAG$^{\ddagger}$ competitor struggles with an MRR@10 of 0.410. This is consistent with its trace generator ignoring system answers, even though the gold page of 48\% of TopiOCQA test questions is named only in an earlier system answer. Furthermore, CMT-RAG$^{\ddagger}$ lacks a mechanism to record withdrawn requests. Consistent with this, it ranks a passage of the rejected request above every gold passage for 25.6\% of TopiOCQA corrections and 36.2\% of self-repairs, against 5.1\% and 7.0\% for \system{}. Operationally, \system{} resolves the dialogue state using a single inference pass through a 3B model per turn, whereas CMT-RAG$^{\ddagger}$ requires approximately three calls to a 32B model per turn. Appendix~\ref{app:qualitative} gives hard cases and an error analysis.

\section{Conclusion}
\label{sec:conclusion}

We introduced \dataset{}, a benchmark of returns and revisions, and \system{}, a memory of topics, facts and operations that a 3B writer revises one turn at a time. \system{} obtains the highest score on all three test sets and is never significantly behind the strongest competitor of any tier. Distillation alone already makes a 3B model a strong rewriter; the memory and the search steps built on it add a further 0.084 MRR@10. Together, they let a 3B model match or outperform prompted methods ten times its size on every test set. Future work can extend the memory to questions that combine several earlier topics.

\section*{Limitations}

\dataset{} covers revisions of a single request. Questions that combine several earlier topics, as in multi-hop questions, were not evaluated. All three corpora are in English and built on Wikipedia. Our TopiOCQA and OR-QuAC search collections are subsets of the full Wikipedia collections, built around the articles used by the benchmark conversations, so our scores cannot be compared with numbers published on the full collections.


\bibliography{refs}

@inproceedings{chiq,
  title={CHIQ: Contextual history enhancement for improving query rewriting in conversational search},
  author={Mo, Fengran and Ghaddar, Abbas and Mao, Kelong and Rezagholizadeh, Mehdi and Chen, Boxing and Liu, Qun and Nie, Jian-Yun},
  booktitle={Proceedings of the 2024 Conference on Empirical Methods in Natural Language Processing},
  pages={2253--2268},
  year={2024}
}

@inproceedings{convgqr,
  title={Convgqr: Generative query reformulation for conversational search},
  author={Mo, Fengran and Mao, Kelong and Zhu, Yutao and Wu, Yihong and Huang, Kaiyu and Nie, Jian-Yun},
  booktitle={Proceedings of the 61st Annual Meeting of the Association for Computational Linguistics (Volume 1: Long Papers)},
  pages={4998--5012},
  year={2023}
}

@inproceedings{chatretriever,
  title={Chatretriever: Adapting large language models for generalized and robust conversational dense retrieval},
  author={Mao, Kelong and Deng, Chenlong and Chen, Haonan and Mo, Fengran and Liu, Zheng and Sakai, Tetsuya and Dou, Zhicheng},
  booktitle={Proceedings of the 2024 Conference on Empirical Methods in Natural Language Processing},
  pages={1227--1240},
  year={2024}
}

@inproceedings{convsearchr1,
  title={Convsearch-r1: Enhancing query reformulation for conversational search with reasoning via reinforcement learning},
  author={Zhu, Changtai and Wang, Siyin and Feng, Ruijun and Song, Kai and Qiu, Xipeng},
  booktitle={Proceedings of the 2025 Conference on Empirical Methods in Natural Language Processing},
  pages={26558--26575},
  year={2025}
}

@inproceedings{canard,
  title={Can you unpack that? learning to rewrite questions-in-context},
  author={Elgohary, Ahmed and Peskov, Denis and Boyd-Graber, Jordan Lee},
  booktitle={Proceedings of the 2019 Conference on Empirical Methods in Natural Language Processing and the 9th International Joint Conference on Natural Language Processing (EMNLP-IJCNLP)},
  pages={5918--5924},
  year={2019}
}

@article{topiocqa,
  title={Topiocqa: Open-domain conversational question answering with topic switching},
  author={Adlakha, Vaibhav and Dhuliawala, Shehzaad and Suleman, Kaheer and de Vries, Harm and Reddy, Siva},
  journal={Transactions of the Association for Computational Linguistics},
  volume={10},
  pages={468--483},
  year={2022},
  publisher={MIT Press One Broadway, 12th Floor, Cambridge, Massachusetts 02142, USA~…}
}

@inproceedings{coral,
  title={CORAL: Benchmarking multi-turn conversational retrieval-augmented generation},
  author={Cheng, Yiruo and Mao, Kelong and Zhao, Ziliang and Dong, Guanting and Qian, Hongjin and Wu, Yongkang and Sakai, Tetsuya and Wen, Ji-Rong and Dou, Zhicheng},
  booktitle={Findings of the Association for Computational Linguistics: NAACL 2025},
  pages={1308--1330},
  year={2025}
}

@inproceedings{orquac,
  title={Open-retrieval conversational question answering},
  author={Qu, Chen and Yang, Liu and Chen, Cen and Qiu, Minghui and Croft, W Bruce and Iyyer, Mohit},
  booktitle={Proceedings of the 43rd International ACM SIGIR conference on research and development in Information Retrieval},
  pages={539--548},
  year={2020}
}

@inproceedings{convdr,
  title={Few-shot conversational dense retrieval},
  author={Yu, Shi and Liu, Zhenghao and Xiong, Chenyan and Feng, Tao and Liu, Zhiyuan},
  booktitle={Proceedings of the 44th International ACM SIGIR Conference on research and development in information retrieval},
  pages={829--838},
  year={2021}
}

@article{t5qr,
  title={Conversational question reformulation via sequence-to-sequence architectures and pretrained language models},
  author={Lin, Sheng-Chieh and Yang, Jheng-Hong and Nogueira, Rodrigo and Tsai, Ming-Feng and Wang, Chuan-Ju and Lin, Jimmy},
  journal={arXiv preprint arXiv:2004.01909},
  year={2020}
}

@article{cmtrag,
  title={CMT-RAG: Complementary Memory Traces for Multi-turn Multi-hop RAG},
  author={Zhou, Lang and Chen, Yingjian and Li, Shuxuan and Lin, Kun-Yu and Zhao, Zhilin},
  journal={arXiv preprint arXiv:2607.26470},
  year={2026}
}

@article{excise,
  title={EXCISE: Query-Side Exclusion for Late-Interaction Retrieval},
  author={Ali, Mohammed and Abdallah, Abdelrahman and Jatowt, Adam},
  journal={arXiv preprint arXiv:2608.05497},
  year={2026}
}

@inproceedings{excluir,
  title     = {{ExcluIR}: Exclusionary Neural Information Retrieval},
  author    = {Zhang, Wenhao and Zhang, Mengqi and Wu, Shiguang and Pei, Jiahuan and Ren, Zhaochun and de Rijke, Maarten and Chen, Zhumin and Ren, Pengjie},
  booktitle = {Proceedings of the AAAI Conference on Artificial Intelligence},
  volume    = {39},
  pages     = {13295--13303},
  year      = {2025}
}

@inproceedings{colbert,
  title={Colbert: Efficient and effective passage search via contextualized late interaction over bert},
  author={Khattab, Omar and Zaharia, Matei},
  booktitle={Proceedings of the 43rd International ACM SIGIR conference on research and development in Information Retrieval},
  pages={39--48},
  year={2020}
}

@misc{reasonmoderncolbert,
  title     = {{Reason-ModernColBERT}},
  author    = {Chaffin, Antoine},
  year      = {2025},
  howpublished = {\url{https://huggingface.co/lightonai/Reason-ModernColBERT}}
}

@article{lora,
  title={Lora: Low-rank adaptation of large language models},
  author={Hu, Edward J and Shen, Yelong and Wallis, Phillip and Allen-Zhu, Zeyuan and Li, Yuanzhi and Wang, Shean and Wang, Lu and Chen, Weizhu},
  journal={arXiv preprint arXiv:2106.09685},
  year={2021}
}

@article{ministral,
  title     = {Ministral 3},
  author    = {Alexander H. Liu and Kartik Khandelwal and Sandeep Subramanian and Victor Jouault and Abhinav Rastogi and Adrien Sadé and Alan Jeffares and Albert Jiang and Alexandre Cahill and Alexandre Gavaudan and Alexandre Sablayrolles and Amélie Héliou and Amos You and Andy Ehrenberg and Andy Lo and Anton Eliseev and Antonia Calvi and Avinash Sooriyarachchi and Baptiste Bout and Baptiste Rozière and Baudouin De Monicault and Clémence Lanfranchi and Corentin Barreau and Cyprien Courtot and Daniele Grattarola and Darius Dabert and Diego de las Casas and Elliot Chane-Sane and Faruk Ahmed and Gabrielle Berrada and Gaëtan Ecrepont and Gauthier Guinet and Georgii Novikov and Guillaume Kunsch and Guillaume Lample and Guillaume Martin and Gunshi Gupta and Jan Ludziejewski and Jason Rute and Joachim Studnia and Jonas Amar and Joséphine Delas and Josselin Somerville Roberts and Karmesh Yadav and Khyathi Chandu and Kush Jain and Laurence Aitchison and Laurent Fainsin and Léonard Blier and Lingxiao Zhao and Louis Martin and Lucile Saulnier and Luyu Gao and Maarten Buyl and Margaret Jennings and Marie Pellat and Mark Prins and Mathieu Poirée and Mathilde Guillaumin and Matthieu Dinot and Matthieu Futeral and Maxime Darrin and Maximilian Augustin and Mia Chiquier and Michel Schimpf and Nathan Grinsztajn and Neha Gupta and Nikhil Raghuraman and Olivier Bousquet and Olivier Duchenne and Patricia Wang and Patrick von Platen and Paul Jacob and Paul Wambergue and Paula Kurylowicz and Pavankumar Reddy Muddireddy and Philomène Chagniot and Pierre Stock and Pravesh Agrawal and Quentin Torroba and Romain Sauvestre and Roman Soletskyi and Rupert Menneer and Sagar Vaze and Samuel Barry and Sanchit Gandhi and Siddhant Waghjale and Siddharth Gandhi and Soham Ghosh and Srijan Mishra and Sumukh Aithal and Szymon Antoniak and Teven Le Scao and Théo Cachet and Theo Simon Sorg and Thibaut Lavril and Thiziri Nait Saada and Thomas Chabal and Thomas Foubert and Thomas Robert and Thomas Wang and Tim Lawson and Tom Bewley and Tom Edwards and Umar Jamil and Umberto Tomasini and Valeriia Nemychnikova and Van Phung and Vincent Maladière and Virgile Richard and Wassim Bouaziz and Wen-Ding Li and William Marshall and Xinghui Li and Xinyu Yang and Yassine El Ouahidi and Yihan Wang and Yunhao Tang and Zaccharie Ramzi},
  journal   = {arXiv preprint arXiv:2601.08584},
  year      = {2026}
}

@article{qwen25,
  title     = {Qwen2.5 Technical Report},
  author    = {{Qwen Team}},
  journal   = {arXiv preprint arXiv:2412.15115},
  year      = {2024}
}

@article{grpo,
  title={Deepseekmath: Pushing the limits of mathematical reasoning in open language models},
  author={Shao, Zhihong and Wang, Peiyi and Zhu, Qihao and Xu, Runxin and Song, Junxiao and Bi, Xiao and Zhang, Haowei and Zhang, Mingchuan and Li, YK and Wu, Yang and others},
  journal={arXiv preprint arXiv:2402.03300},
  year={2024}
}

@article{ance,
  title={Approximate nearest neighbor negative contrastive learning for dense text retrieval},
  author={Xiong, Lee and Xiong, Chenyan and Li, Ye and Tang, Kwok-Fung and Liu, Jialin and Bennett, Paul and Ahmed, Junaid and Overwijk, Arnold},
  journal={arXiv preprint arXiv:2007.00808},
  year={2020}
}

@inproceedings{recor,
  title={Recor: Reasoning-focused multi-turn conversational retrieval benchmark},
  author={Ali, Mohammed and Abdallah, Abdelrahman and Agarwal, Amit and Patel, Hitesh Laxmichand and Jatowt, Adam},
  booktitle={Findings of the Association for Computational Linguistics: ACL 2026},
  pages={2688--2723},
  year={2026}
}

@article{mem0,
  title={Mem0: Building production-ready ai agents with scalable long-term memory},
  author={Chhikara, Prateek and Khant, Dev and Aryan, Saket and Singh, Taranjeet and Yadav, Deshraj},
  journal={arXiv preprint arXiv:2504.19413},
  year={2025}
}

@article{zep,
  title={Zep: a temporal knowledge graph architecture for agent memory},
  author={Rasmussen, Preston and Paliychuk, Pavlo and Beauvais, Travis and Ryan, Jack and Chalef, Daniel},
  journal={arXiv preprint arXiv:2501.13956},
  year={2025}
}

@inproceedings{contextagent,
  title={Context-agent: Dynamic discourse trees for non-linear dialogue},
  author={Hu, Junan and Guo, Shudan and Liu, Wenqi and Yin, Jianhua and Wei, Yinwei},
  booktitle={Findings of the Association for Computational Linguistics: ACL 2026},
  pages={29449--29462},
  year={2026}
}

@article{schegloff1977,
  title={The preference for self-correction in the organization of repair in conversation},
  author={Schegloff, Emanuel A and Jefferson, Gail and Sacks, Harvey},
  journal={Language},
  volume={53},
  number={2},
  pages={361--382},
  year={1977},
  publisher={Cambridge University Press}
}

@inproceedings{haconvdr,
  title={History-aware conversational dense retrieval},
  author={Mo, Fengran and Qu, Chen and Mao, Kelong and Zhu, Tianyu and Su, Zhan and Huang, Kaiyu and Nie, Jian-Yun},
  booktitle={Findings of the Association for Computational Linguistics: ACL 2024},
  pages={13366--13378},
  year={2024}
}

@article{longmemeval,
  title={Longmemeval: Benchmarking chat assistants on long-term interactive memory},
  author={Wu, Di and Wang, Hongwei and Yu, Wenhao and Zhang, Yuwei and Chang, Kai-Wei and Yu, Dong},
  journal={arXiv preprint arXiv:2410.10813},
  year={2024}
}

@inproceedings{beam,
  title={Beyond a million tokens: Benchmarking and enhancing long-term memory in LLMs},
  author={Tavakoli, Mohammad and Salemi, Alireza and Ye, Carrie and Abdalla, Mohamed and Zamani, Hamed and Mitchell, J},
  booktitle={International Conference on Learning Representations},
  volume={2026},
  pages={133343--133429},
  year={2026}
}

@inproceedings{locomo,
  title={Evaluating very long-term conversational memory of llm agents},
  author={Maharana, Adyasha and Lee, Dong-Ho and Tulyakov, Sergey and Bansal, Mohit and Barbieri, Francesco and Fang, Yuwei},
  booktitle={Proceedings of the 62nd Annual Meeting of the Association for Computational Linguistics (Volume 1: Long Papers)},
  pages={13851--13870},
  year={2024}
}

@article{memgpt,
  title={Memgpt: Towards llms as operating systems},
  author={Packer, Charles and Wooders, Sarah and Lin, Kevin and Fang, Vivian and Patil, Shishir G and Stoica, Ion and Gonzalez, Joseph E},
  journal={arXiv preprint arXiv:2310.08560},
  year={2023}
}

@article{sinha2026context,
  title={Beyond Continuity: Challenges of Context Switching in Multi-Turn Dialogue with LLMs},
  author={Sinha, Aditya and Steck, Harald and Ostuni, Vito and Rinaldi, Matteo},
  journal={arXiv preprint arXiv:2605.09268},
  year={2026}
}

@inproceedings{multiwoz,
  title={Multiwoz-a large-scale multi-domain wizard-of-oz dataset for task-oriented dialogue modelling},
  author={Budzianowski, Pawe{\l} and Wen, Tsung-Hsien and Tseng, Bo-Hsiang and Casanueva, I{\~n}igo and Ultes, Stefan and Ramadan, Osman and Gasic, Milica},
  booktitle={Proceedings of the 2018 conference on empirical methods in natural language processing},
  pages={5016--5026},
  year={2018}
}

@inproceedings{trade,
  title={Transferable multi-domain state generator for task-oriented dialogue systems},
  author={Wu, Chien-Sheng and Madotto, Andrea and Hosseini-Asl, Ehsan and Xiong, Caiming and Socher, Richard and Fung, Pascale},
  booktitle={Proceedings of the 57th annual meeting of the association for computational linguistics},
  pages={808--819},
  year={2019}
}

@inproceedings{memorybank,
  title={{MemoryBank}: Enhancing Large Language Models with Long-Term Memory},
  author={Zhong, Wanjun and Guo, Lianghong and Gao, Qiqi and Ye, He and Wang, Yanlin},
  booktitle={Proceedings of the AAAI Conference on Artificial Intelligence},
  volume={38},
  pages={19724--19731},
  year={2024}
}

@article{amem,
  title={A-mem: Agentic memory for llm agents},
  author={Xu, Wujiang and Liang, Zujie and Mei, Kai and Gao, Hang and Tan, Juntao and Zhang, Yongfeng},
  journal={Advances in Neural Information Processing Systems},
  volume={38},
  pages={17577--17604},
  year={2026}
}

@inproceedings{quac,
  title={QuAC: Question answering in context},
  author={Choi, Eunsol and He, He and Iyyer, Mohit and Yatskar, Mark and Yih, Wen-tau and Choi, Yejin and Liang, Percy and Zettlemoyer, Luke},
  booktitle={Proceedings of the 2018 conference on empirical methods in natural language processing},
  pages={2174--2184},
  year={2018}
}

@inproceedings{msc,
  title={Beyond goldfish memory: Long-term open-domain conversation},
  author={Xu, Jing and Szlam, Arthur and Weston, Jason},
  booktitle={Proceedings of the 60th annual meeting of the association for computational linguistics (volume 1: long papers)},
  pages={5180--5197},
  year={2022}
}

@inproceedings{bright,
  title={Bright: A realistic and challenging benchmark for reasoning-intensive retrieval},
  author={Su, Hongjin and Yen, Howard and Xia, Mengzhou and Shi, Weijia and Muennighoff, Niklas and Wang, Han-yu and Haisu, Liu and Shi, Quan and Siegel, Zachary and Tang, Michael and others},
  booktitle={International Conference on Learning Representations},
  volume={2025},
  pages={48941--48991},
  year={2025}
}

@inproceedings{hirschberg2001,
  title={Identifying user corrections automatically in spoken dialogue systems},
  author={Hirschberg, Julia and Litman, Diane and Swerts, Marc},
  booktitle={Second Meeting of the North American Chapter of the Association for Computational Linguistics},
  year={2001}
}

@article{litman2006,
  title={Characterizing and predicting corrections in spoken dialogue systems},
  author={Litman, Diane and Hirschberg, Julia and Swerts, Marc},
  journal={Computational linguistics},
  volume={32},
  number={3},
  pages={417--438},
  year={2006},
  publisher={MIT Press One Rogers Street, Cambridge, MA 02142-1209, USA journals-info~…}
}

\clearpage
\appendix
\raggedbottom
\setcounter{dbltopnumber}{3}\renewcommand{\dbltopfraction}{0.9}
\section{Extended Related Work}
\label{app:related}

\begin{table*}[t]
\centering
\footnotesize
\setlength{\tabcolsep}{4.5pt}
\begin{tabular}{@{}l c cc cc c@{}}
\toprule
\multirow{2}{*}{\textbf{Method}} & \textit{History} & \multicolumn{2}{c}{\textit{Memory}} & \multicolumn{2}{c}{\textit{Search}} & \textit{Evaluation} \\
\cmidrule(lr){2-2}\cmidrule(lr){3-4}\cmidrule(lr){5-6}\cmidrule(lr){7-7}
 & \makecell{Read\\once} & Kept & Revisable & \makecell{No gen.\\answer} & \makecell{Frozen\\retr.} & \makecell{Passage\\ranking} \\
\midrule
T5QR \citep{t5qr}                    & \xmark & \xmark & \xmark & \cmark & \cmark & \cmark \\
ConvGQR \citep{convgqr}              & \xmark & \xmark & \xmark & \xmark & \cmark & \cmark \\
ConvSearch-R1 \citep{convsearchr1}   & \xmark & \xmark & \xmark & $\sim$ & \cmark & \cmark \\
CHIQ \citep{chiq}                    & \xmark & \xmark & \xmark & \xmark & \cmark & \cmark \\
ChatRetriever \citep{chatretriever}  & \xmark & \xmark & \xmark & \cmark & \xmark & \cmark \\
CMT-RAG \citep{cmtrag}               & \cmark & \cmark & \xmark & \xmark & \cmark & \xmark \\
\midrule
\textbf{\system{}} (ours)            & \cmark & \cmark & \cmark & \cmark & \cmark & \cmark \\
\bottomrule
\end{tabular}
\par\smallskip
\parbox{\textwidth}{\footnotesize \cmark{} yes, $\sim$ partly, \xmark{} no. Read once: each turn is processed once, without re-reading the history. Kept: a memory is carried across turns. Revisable: a later turn can change the status of stored content. No gen.\ answer: the query does not depend on a generated answer. Frozen retr.: the retriever is used as released. -FT variants share their method's row, and LLM-Rewrite is a prompted rewrite like CHIQ.}
\caption{Conversational search methods along the dimensions that matter for revisions.}
\label{tab:comparison}
\end{table*}

This appendix extends \S\ref{sec:related} and follows the columns of Table~\ref{tab:comparison}.

\paragraph{Conversational query rewriting.}
The standard approach to searching in a conversation is to rewrite the last question into a standalone query and submit it to an ad hoc retriever. Trained rewriters learn this from human rewrites, as T5QR \citep{t5qr} does on CANARD \citep{canard}. ConvGQR \citep{convgqr} appends a generated answer to the rewrite and, during training, pulls the rewriter's representation of the conversation towards the retriever's embedding of the gold passage. ConvSearch-R1 \citep{convsearchr1} does not use human rewrites. It first trains on its own rewrites that rank the gold passage first, and then optimises the rewriter by reinforcement learning with a reward derived from the rank of the gold passage. Prompted rewriters obtain the rewrite from a large language model. CHIQ \citep{chiq} first enhances the history with an open language model. It disambiguates the new question, expands the last system response, generates a pseudo-response, keeps only the last turn after a topic switch and summarises the history, and it then rewrites the question from the enhanced history. All of these methods re-read the whole history at every turn and output a single query. None records that an earlier request was withdrawn, so a revision must be recognised anew, from the raw text, whenever a query is formed.

\paragraph{Conversational dense retrieval.}
A second line removes the rewriting step and encodes the conversation directly as the query. ConvDR \citep{convdr} encodes the last question together with all earlier questions and trains this encoder to imitate the embedding an ad hoc retriever assigns to the human rewrite. ChatRetriever \citep{chatretriever} adapts a large language model (Qwen-7B-Chat) for this purpose, combining contrastive learning with session-masked instruction tuning. HAConvDR \citep{haconvdr} judges an earlier turn useful when adding it to the current question improves retrieval, and uses these judgments to train a retriever on a denoised history. These retrievers also read the earlier turns at every turn. Whether a topic is still wanted must be decided inside a single query vector, where the decision cannot be inspected.

\paragraph{Memory for multi-turn RAG.}
CMT-RAG \citep{cmtrag} stores, for every turn, a trace of sub-questions with their dependencies on earlier turns, the passages retrieved for them and the reader's answer, and recovers these traces in later turns. Its trace generator is a trained state-space model, and the method is evaluated on multi-hop question answering on its MuMu-QA benchmark. Our memory instead records the topics of the conversation and what each turn did to them, while CMT-RAG keeps a growing record of what was asked. The exclusion step of our search follows EXCISE \citep{excise}, which identifies the excluded topic outside the index and demotes matching candidates over a frozen index; exclusionary queries of this kind are studied in ExcluIR \citep{excluir}.

\paragraph{Memory for chat agents.}
MemGPT \citep{memgpt} moves information between the context window of a language model and external storage, so that an agent can continue a conversation over many sessions. MemoryBank \citep{memorybank} stores memories of past conversations and lets them fade with time unless they are recalled. Mem0 \citep{mem0} extracts facts from each exchange and decides for each one whether to add it, update or delete an existing memory, or change nothing. Zep \citep{zep} stores facts as edges of a temporal knowledge graph and marks an edge as no longer valid when new information contradicts it. A-Mem \citep{amem} keeps memories as linked notes, and a new note can update the descriptions of related older ones. Context-Agent \citep{contextagent} organises a dialogue into trees of topics and branches and builds the context of each response from the active branch and summaries of the others. These systems are evaluated through the responses of a generator, on benchmarks that test whether facts from long conversations are recalled, as in LoCoMo \citep{locomo}, and kept up to date as new information arrives, as in LongMemEval \citep{longmemeval} and BEAM \citep{beam}. \citet{sinha2026context} build dialogues from TopiOCQA and MSC \citep{msc} and test whether language models notice when the user moves to a new topic and leave the earlier context behind. Dialogue state tracking follows the values a user gives for a fixed set of slots in task-oriented dialogue, such as the area or price range of a restaurant, and a later turn can change a value \citep{multiwoz,trade}.

\paragraph{Benchmarks.}
TopiOCQA \citep{topiocqa} collects information-seeking conversations over Wikipedia in which the questioner may switch to a related topic; from the sixth turn onwards, questioners frequently go back one or two topics, and we use these returns as natural questions. CORAL \citep{coral} is built automatically from Wikipedia: section titles of one or two related pages are sampled as a path and converted into conversational questions by GPT-4, and the passages come from the web pages the articles cite. OR-QuAC \citep{orquac} moves QuAC \citep{quac} to an open-retrieval setting over Wikipedia; each conversation concerns one article, and its answers lie in the same section of that article. We use the three because they test different conditions: TopiOCQA has human topic switches with a topic label for every turn, from which natural returns can be found, and several compared methods report results on it; OR-QuAC has human questions over passages with page titles; and CORAL has passages from cited web pages without page titles, so it tests \system{} where the title-based search steps cannot help, and it also supplies training conversations. RECOR \citep{recor} builds 707 multi-turn conversations in eleven domains from BRIGHT \citep{bright} and StackExchange queries to test reasoning-intensive retrieval. None of these four papers describes turns in which the user corrects or revises an earlier request, so the corrections and revisions of \S\ref{sec:benchmark} had to be written. \dataset{} adds these questions, labels every question with its tier, and provides every written question with its stale passages.

\section{Benchmark Construction}
\label{app:benchmark}

\begin{table}[t]
\centering
\footnotesize
\begin{tabular}{@{}lrrr@{}}
\toprule
\textbf{Tier} & \textbf{TopiOCQA} & \textbf{CORAL} & \textbf{OR-QuAC} \\
\midrule
Return-1 & 1{,}028 & 167 & 38 \\
Return-2+ & 572 & 188 & 55 \\
Correction & 759 & 262 & 218 \\
Follow-up & 300 & -- & -- \\
Self-repair & 608 & -- & -- \\
Value revision & 233 & -- & 1 \\
\midrule
Total & 3{,}500 & 617 & 312 \\
Conversations & 1{,}196 & 268 & 219 \\
\bottomrule
\end{tabular}
\caption{Test questions per tier and corpus. --: tier not built for this corpus.}
\label{tab:bench_stats}
\end{table}

Table~\ref{tab:bench_stats} gives the number of test questions per tier and corpus. Returns come from natural TopiOCQA turns and spliced conversations, and 104 Return-2+ questions are written; corrections come from built revisions and spliced conversations, follow-ups and self-repairs from built revisions, and value revisions as described below.

\paragraph{Natural returns.}
A TopiOCQA turn with an answer is a return when its topic label differs from that of the previous turn and appeared earlier in the conversation. We keep only returns whose question refers to the target by a pronoun or a description. Its depth, the number of other topics discussed since the target, decides between Return-1 (depth one) and Return-2+.

\paragraph{Built revisions.}
We select two topics of the same kind from one TopiOCQA conversation and truncate the conversation after the last turn about either of them. We then add a question that applies to both, answer it for the wrong topic, and append a correction, a self-repair, or a correction followed by a new question. Each item is built for both topics with the same history. The gold passage belongs to the intended topic and the stale passages to the other one.

\paragraph{Written returns and value revisions.}
For 104 Return-2+ questions we append one turn to a TopiOCQA conversation with at least four topics, and this turn returns to a middle topic by describing it. A value revision follows a turn whose page has a sibling that differs by a number, such as a year or an edition, with a short switch such as ``sorry, i meant 2, not 1''. Its gold passage belongs to the sibling and its stale passage to the original page.

\paragraph{Spliced conversations.}
A spliced item keeps the opening of a conversation about a topic $E$ and appends digressions about two further topics, each linked to $E$, confusable with it, or taken from another conversation. The final turn targets one of the three topics: it is either the source's own next turn about $E$ or a new question about an unseen passage of the target, written as a correction or as a return by pronoun or description. A return to $E$ is Return-2+, a return to the first digression Return-1, and the passages of the other two topics are stale. On TopiOCQA and CORAL a correction describes the intended topic through an earlier fact; on OR-QuAC the digressions are the opening turns of two other OR-QuAC conversations, and the wordings start from templates.

\paragraph{Gold passages and collections.}
Every gold passage is present in its collection, no stale passage is gold, and no mined passage contains an answer from the same conversation. The CORAL collection is the official corpus (162{,}650 passages). The TopiOCQA collection (434{,}179 passages) combines background passages sampled from TopiOCQA's Wikipedia collection, all passages of the articles the benchmark touches, and BM25 hard negatives mined from the full Wikipedia for each final question; the OR-QuAC collection (73{,}992 passages) is built in the same way from the official OR-QuAC collection.

\paragraph{Source splits.}
Of the TopiOCQA test questions, 2{,}839 (81\%) come from conversations in TopiOCQA's official training file, which favours released models trained on that file.

\paragraph{Training and development data.}
The 1{,}790 training questions comprise 839 from TopiOCQA training conversations, built in the same ways as the test set, and 951 from CORAL's training split (714 spliced, 237 prefixes of natural CORAL conversations); they come from 769 conversations. The six development sets are one TopiOCQA set (431 questions of all six tiers), three CORAL sets (300 questions from natural conversations, 450 corrections and returns, and 300 returns) and two OR-QuAC sets (218 corrections and returns, and 84 value revisions).

\section{System Details}
\label{app:system}

Algorithm~\ref{alg:memrevise} summarises how \system{} processes one conversation. The following paragraphs give the details omitted from \S\ref{sec:method}.

\begin{algorithm}[t]
\caption{\system{} on one conversation}
\label{alg:memrevise}
\small
\begin{algorithmic}[1]
\Require turns $(q_1,a_1),\ldots,(q_{n-1},a_{n-1})$, final question $q_n$ ($a_n$ unknown), writer $\pi$, collection $\mathcal{P}$
\State $M \gets \emptyset$ \Comment{memory: nodes and topic statuses}
\For{$t = 1$ \textbf{to} $n$}
  \State $x \gets \textsc{Render}(M) \oplus q_{t-1} \oplus (q_t, a_t)$
  \State $\nu_t \gets \pi(x)$ \Comment{operations, topic, facts; query if $t=n$}
  \If{$1<t<n$ \textbf{and} $\mathcal{P}$ has page titles}
    \State $\nu_t \gets \textsc{RepairTopic}(\nu_t, q_t, a_t)$ \Comment{\S\ref{sec:repair}}
  \EndIf
  \State $M \gets M \cup \{\nu_t\}$ \Comment{statuses replayed from operations}
\EndFor
\If{$q_n$ only replaces a number (number rule)}
  \State $s \gets z\big(S(q^{s\prime},\cdot)\big)$ + title prior if titled \Comment{new number}
\Else
  \State $s \gets z\big(0.67\,S(q^{s},\cdot) + 0.33\,S(q^{t},\cdot)\big)$ \Comment{Eq.~\ref{eq:score}}
  \If{$\mathcal{P}$ has page titles}
    \State add the title prior ($+1.5$) and the facts vote ($+1.0$)
    \If{$\nu_n$ rejects a topic whose page shares its name} \State apply the exclusion step ($\pm1.5$) \EndIf
  \EndIf
\EndIf
\State \Return passages of $\mathcal{P}$ ranked by $s$
\end{algorithmic}
\end{algorithm}

\paragraph{Input and output.}
The rendered memory lists every topic in order of first mention with its kind, the turns that discussed it, its status and its stored facts, followed by a log with one line per turn. Statuses are replayed as in \S\ref{sec:memory}: \textsc{Add}, \textsc{Refine}, \textsc{Revise} and \textsc{Reactivate} make their topic current and move the previously current topic to earlier, \textsc{Abandon} moves its topic to dropped, and the operations of one turn are applied in the order the writer lists them; a dropped topic becomes current again with any later operation on it, usually \textsc{Reactivate}. \textsc{Reactivate} also restores a name first given in an earlier answer, as for Neil Armstrong in Figure~\ref{fig:prompt_example}. A node may list several topics, and the first is its topic. A switch to a different entity with the same name is written as \textsc{Abandon} on the old topic and \textsc{Revise} to the new one. The final line of the input indicates whether a query is requested, which is the case only for the final turn. The writer responds with one JSON object in the order described in \S\ref{sec:student}. When the user refers to a different entity with the same name as a stored topic, the writer keeps the name and appends the user's description, as in ``Jayson Blair (actor)''. History answers are cut at 600 characters, as for every method. Inputs are capped at 16{,}384 tokens, which no input came close to, and outputs at 800 tokens, decoded greedily. Figure~\ref{fig:prompt} gives the full system prompt and Figure~\ref{fig:prompt_example} the input and output for turn 4 of the conversation in Figure~\ref{fig:example}, searched as the final turn.

\begin{figure*}[p]
\begin{tcolorbox}[colback=slBg, colframe=slLine, boxrule=0.5pt, arc=2pt, left=4pt, right=4pt, top=3pt, bottom=3pt,
  title={\small\textbf{System prompt of the writer}}, coltitle=slInk, colbacktitle=slBub]
\fontsize{6.4}{7.6}\selectfont\ttfamily\raggedright\parindent=0pt
You update a conversation MEMORY one turn at a time. You never see the conversation itself: you see the MEMORY built from the earlier turns (every topic in the order it came up, its kind, the turns it was discussed in, its status, the facts stored for it, and a log of what each earlier turn did), the LAST QUESTION kept in memory, and the NEW TURN (its question, and its answer when the answer is already known). The NEW TURN is the CURRENT turn below. Analyze ONLY the CURRENT turn (use the MEMORY to resolve references and dependencies; earlier turn numbers are the ones shown in the memory). Output ONE JSON object:\par
\{\par
\hspace*{1.0em}\textquotedbl{}depends\_on\textquotedbl{}: [\textless{}prior turn numbers this turn relies on; [] if fully self-{}contained\textgreater{}],\par
\hspace*{1.0em}\textquotedbl{}contradicts\textquotedbl{}: [\textquotedbl{}\textless{}prior entity name(s) this turn CORRECTS/REPLACES; [] otherwise\textgreater{}\textquotedbl{}],\par
\hspace*{1.0em}\textquotedbl{}ops\textquotedbl{}: [\{\textquotedbl{}op\textquotedbl{}: \textquotedbl{}\textless{}ADD\textbar{}REFINE\textbar{}REVISE\textbar{}ABANDON\textbar{}REACTIVATE\textbar{}RETRACT\textgreater{}\textquotedbl{}, \textquotedbl{}entity\textquotedbl{}: \textquotedbl{}\textless{}canonical name\textgreater{}\textquotedbl{}, \textquotedbl{}detail\textquotedbl{}: \textquotedbl{}\textless{}optional short phrase\textgreater{}\textquotedbl{}\}],\par
\hspace*{1.0em}\textquotedbl{}asked\textquotedbl{}: [\textquotedbl{}\textless{}canonical full name(s) of the entity/entities the CURRENT turn is ABOUT; resolve every pronoun/ellipsis; usually 1, up to 3 for a multi-{}topic turn\textgreater{}\textquotedbl{}],\par
\hspace*{1.0em}\textquotedbl{}query\textquotedbl{}: \textquotedbl{}\textless{}ONLY when the prompt ends with WRITE THE QUERY: yes -{}-{} ONE search query for this turn; otherwise leave this field out\textgreater{}\textquotedbl{},\par
\hspace*{1.0em}\textquotedbl{}facts\textquotedbl{}: [\{\textquotedbl{}entity\textquotedbl{}: \textquotedbl{}\textless{}canonical full name this fact is about\textgreater{}\textquotedbl{}, \textquotedbl{}fact\textquotedbl{}: \textquotedbl{}\textless{}ONE short grounded SENTENCE the CURRENT turn\textquotesingle{}s question+answer establishes\textgreater{}\textquotedbl{}\}],\par
\hspace*{1.0em}\textquotedbl{}types\textquotedbl{}: \{\textquotedbl{}\textless{}entity name that is NEW in this turn\textgreater{}\textquotedbl{}: \textquotedbl{}\textless{}one of [\textquotesingle{}person\textquotesingle{}, \textquotesingle{}work\textquotesingle{}, \textquotesingle{}place\textquotesingle{}, \textquotesingle{}organization\textquotesingle{}, \textquotesingle{}event\textquotesingle{}, \textquotesingle{}date\textquotesingle{}, \textquotesingle{}concept\textquotesingle{}, \textquotesingle{}object\textquotesingle{}, \textquotesingle{}other\textquotesingle{}]\textgreater{}\textquotedbl{}\}\par
\}\par
\par\vspace{1.2pt}
RULES (read carefully):\par
-{} facts: extract EVERY salient fact the CURRENT turn\textquotesingle{}s Q+A establishes \textemdash{} one NATURAL SENTENCE per\par
\hspace*{1.0em}fact (2-{}5 typical), NOT a single summary. PRESERVE the exact relation and terms from the answer\par
\hspace*{1.0em}(\textquotedbl{}Dunkirk was released on DVD on 18 December 2017\textquotedbl{}, NOT \textquotedbl{}Dunkirk has-{}property 18 December 2017\textquotedbl{};\par
\hspace*{1.0em}keep \textquotedbl{}shot in\textquotedbl{}, \textquotedbl{}is about\textquotedbl{}, \textquotedbl{}krill\textquotedbl{} \textemdash{} never crush them into a generic predicate). Full canonical\par
\hspace*{1.0em}names, no pronouns.\par
-{} depends\_on (LOAD-{}BEARING \textemdash{} be precise): the prior turn numbers that introduce the entity or the\par
\hspace*{1.0em}fact THIS turn presupposes. ALWAYS include the turn that first named the entity in \textquotedbl{}asked\textquotedbl{}, PLUS\par
\hspace*{1.0em}any turn whose answer this question builds on. Use [] ONLY if the turn is fully self-{}contained\par
\hspace*{1.0em}(introduces its own subject and needs no earlier context).\par
-{} asked: the resolved TARGET (what a search must be ABOUT), never a pronoun.\par
-{} ENTITY NAMES: use the SHORTEST canonical common name (\textquotedbl{}Urk\textquotedbl{}, \textquotedbl{}Dunkirk\textquotedbl{}, \textquotedbl{}Mercury\textquotedbl{}), NOT a long\par
\hspace*{1.0em}legal/disambiguated title (\textquotedbl{}High and Low Lordship of Urk and Emmeloord\textquotedbl{}) unless disambiguation\par
\hspace*{1.0em}truly requires it. Be consistent across turns so the same entity links by exact name.\par
-{} contradicts: emit ONLY when the turn retracts/corrects a prior premise (\textquotedbl{}no, I meant X\textquotedbl{});\par
\hspace*{1.0em}otherwise [].\par
-{} ops (what this turn DOES to the conversation\textquotesingle{}s memory; most turns emit exactly one):\par
\hspace*{2.0em}ADD        a new topic entity enters the conversation\par
\hspace*{2.0em}REFINE     stays on the current entity, narrower facet (typical follow-{}up)\par
\hspace*{2.0em}REVISE     a stated constraint/value about the entity is replaced with a new one\par
\hspace*{2.0em}ABANDON    the user walks away from an entity (\textquotedbl{}not that one\textquotedbl{}, \textquotedbl{}forget the X thing\textquotedbl{})\par
\hspace*{2.0em}REACTIVATE the user returns to an entity discussed EARLIER (\textquotedbl{}the one I asked about before\textquotedbl{})\par
\hspace*{2.0em}RETRACT    a prior premise is declared factually WRONG (stronger than ABANDON; also\par
\hspace*{7.5em}list that entity in contradicts)\par
\hspace*{1.0em}A correction turn typically emits ABANDON (the rejected topic(s)) + REACTIVATE (the\par
\hspace*{1.0em}restored topic) TOGETHER. Resolve vague references (\textquotedbl{}the one from before\textquotedbl{}) to the\par
\hspace*{1.0em}canonical entity name using the MEMORY.\par
-{} types: optional; tag entities you are confident about. Use ONLY the current turn\textquotesingle{}s Q+A; never\par
\hspace*{1.0em}invent facts and never use later turns. Output ONLY the JSON.\par
NAMES: an entity name is the NAME of a thing (a person, work, place, organization, concept). NEVER use the user\textquotesingle{}s\par
question or a search string as a name, and never a long description when the thing has a name in the MEMORY or the NEW TURN.\par
RETURNS: the user may go back to ANY earlier topic, often by DESCRIBING it (\textquotedbl{}going back to that Indian state\textquotedbl{},\par
\textquotedbl{}the singer of that song\textquotedbl{}, \textquotedbl{}the first city mentioned\textquotedbl{}). Resolve the description to that topic\textquotesingle{}s name and REACTIVATE it.\par
TWO ENTITIES: when the turn is about two things (a person AND their company, a singer AND the song), list both in\par
\textquotedbl{}asked\textquotedbl{}, the one the answer passage is most likely filed under FIRST.\par
\par\vspace{1.2pt}
ORDER: write the fields in the order shown above: first what the turn builds on and what the user DOES (depends\_on, contradicts, ops), then the topic it is about (asked), then the query when it is wanted.\par
SAME NAME: when the user means a DIFFERENT thing that has the same name as a topic in the MEMORY (\textquotesingle{}i meant the actor, not the journalist\textquotesingle{}), keep the name and add the user\textquotesingle{}s description in brackets: \textquotesingle{}Jayson Blair (actor)\textquotesingle{}; ABANDON the old topic and REVISE to the new one. Brackets belong to NAMES only: in the TURN LOG an action\textquotesingle{}s detail is written after a dash.\par
TYPES: give a type only for an entity that is NEW in this turn; the MEMORY already shows the type of the others.\par
QUERY: when the prompt ends with \textquotesingle{}WRITE THE QUERY: yes\textquotesingle{}, add \textquotedbl{}query\textquotedbl{}: ONE natural search question about the topic in \textquotedbl{}asked\textquotedbl{}. Keep the user\textquotesingle{}s question whole, with every pronoun and description replaced by the topic\textquotesingle{}s exact name. If the turn only corrects or points (\textquotesingle{}i meant X, not Y\textquotesingle{}, \textquotesingle{}no, the director\textquotesingle{}) and asks nothing, ask the LAST QUESTION IN MEMORY about the topic. If the turn refers to something inside a stored fact (a list member, \textquotesingle{}the second actor\textquotesingle{}, \textquotesingle{}his wife\textquotesingle{}), name that thing from the facts. Never mention a rejected topic. When the prompt ends with \textquotesingle{}WRITE THE QUERY: no\textquotesingle{}, leave \textquotedbl{}query\textquotedbl{} out.\par
\end{tcolorbox}
\caption{The writer's system prompt, verbatim. The same prompt is used in training and at test time.}
\label{fig:prompt}
\end{figure*}

\begin{figure*}[t]
\begin{minipage}[t]{0.6\textwidth}
\begin{tcolorbox}[colback=slBg, colframe=slLine, boxrule=0.5pt, arc=2pt, left=4pt, right=4pt, top=3pt, bottom=3pt,
  title={\small\textbf{Input (user message) at turn 4}}, coltitle=slInk, colbacktitle=slBub]
\fontsize{6.4}{7.6}\selectfont\ttfamily\raggedright\parindent=0pt
MEMORY\par
1. Apollo 11, a event \textbar{} turns 1, 2 \textbar{} EARLIER\par
\hspace*{2.5em}-{} Apollo 11 landed on the Moon on July 20, 1969.\par
\hspace*{2.5em}-{} Neil Armstrong walked on the Moon first, then Buzz Aldrin.\par
2. Buzz Aldrin, a person \textbar{} turn 3 \textbar{} CURRENT\par
\hspace*{2.5em}-{} Buzz Aldrin grew up in Montclair, New Jersey.\par
\par\vspace{1.2pt}
TURN LOG\par
-{} turn 1: ADD Apollo 11\par
-{} turn 2: REFINE Apollo 11 \textemdash{} Moon landing \textbar{} builds on turn 1\par
-{} turn 3: ADD Buzz Aldrin \textbar{} builds on turn 2\par
\par\vspace{1.2pt}
LAST QUESTION IN MEMORY: where did he grow up?\par
\par\vspace{1.2pt}
NEW TURN (turn 4)\par
Q: no, i meant the second guy... wait, no, the first one. sorry\par
A: (not known yet)\par
\par\vspace{1.2pt}
WRITE THE QUERY: yes\par
\end{tcolorbox}
\end{minipage}\hfill
\begin{minipage}[t]{0.38\textwidth}
\begin{tcolorbox}[colback=white, colframe=pBlue, boxrule=0.5pt, arc=2pt, left=4pt, right=4pt, top=3pt, bottom=3pt,
  title={\small\textbf{Output of the writer}}, coltitle=white, colbacktitle=pBlue]
\fontsize{6.4}{7.6}\selectfont\ttfamily\raggedright\parindent=0pt
\{\textquotedbl{}depends\_on\textquotedbl{}: [2, 3],\par
\hspace*{0.5em}\textquotedbl{}contradicts\textquotedbl{}: [\textquotedbl{}Buzz Aldrin\textquotedbl{}],\par
\hspace*{0.5em}\textquotedbl{}ops\textquotedbl{}: [\{\textquotedbl{}op\textquotedbl{}: \textquotedbl{}ABANDON\textquotedbl{},\par
\hspace*{1.5em}\textquotedbl{}entity\textquotedbl{}: \textquotedbl{}Buzz Aldrin\textquotedbl{}, \textquotedbl{}detail\textquotedbl{}: \textquotedbl{}\textquotedbl{}\},\par
\hspace*{1em}\{\textquotedbl{}op\textquotedbl{}: \textquotedbl{}REACTIVATE\textquotedbl{},\par
\hspace*{1.5em}\textquotedbl{}entity\textquotedbl{}: \textquotedbl{}Neil Armstrong\textquotedbl{}, \textquotedbl{}detail\textquotedbl{}: \textquotedbl{}\textquotedbl{}\}],\par
\hspace*{0.5em}\textquotedbl{}asked\textquotedbl{}: [\textquotedbl{}Neil Armstrong\textquotedbl{}],\par
\hspace*{0.5em}\textquotedbl{}query\textquotedbl{}: \textquotedbl{}where did Neil Armstrong grow up?\textquotedbl{},\par
\hspace*{0.5em}\textquotedbl{}facts\textquotedbl{}: [],\par
\hspace*{0.5em}\textquotedbl{}types\textquotedbl{}: \{\textquotedbl{}Neil Armstrong\textquotedbl{}: \textquotedbl{}person\textquotedbl{}\}\}\par
\end{tcolorbox}
\end{minipage}
\caption{The writer's input and output at turn 4 of the conversation in Figure~\ref{fig:example}, searched as the final turn. Input: the run's prompt, verbatim, with the memory of turns 1--3 as intended; the answer to turn 4 is not known. Output: the writer's output, verbatim (seed 3). The writer rejects Buzz Aldrin, returns to Neil Armstrong and asks the open question.}
\label{fig:prompt_example}
\end{figure*}

\paragraph{Training.}
The writer is trained on 22{,}163 examples from 9{,}916 turns of the TopiOCQA and CORAL training conversations: history turns, final turns, and copies of the history turns that change the memory, with their answer removed, so that the writer also learns to decide without an answer, as at the final turn. Most turns appear twice, once with the teacher memory and once with the memory of an auxiliary writer trained on the other half of the conversations. The competitors are trained on final questions only. We train LoRA adapters (rank 16, $\alpha = 32$, dropout 0.05) on the attention projections of Ministral-3-3B for three epochs, with a learning rate of $5\times10^{-5}$ chosen on development data and an effective batch size of 12; the checkpoint is chosen by the held-out loss on the fields written before the query.

\paragraph{Memory-free writer.}
The memory-free writer of Table~\ref{tab:control} uses the writer's base model and LoRA settings, its learning rate, effective batch size and number of epochs, and three seeds, and its checkpoint is chosen by held-out loss. It reads the raw conversation, with answers cut at 600 characters as for every competitor, and is trained on the writer's final-turn target queries, each turn once. It has no template part, no priors, no exclusion step and no topic repair, and its queries are scored with the same frozen retriever as every other method.

\paragraph{Search rules.}
From the second turn on, topic repair runs two searches, one with the turn's question and answer and one with the stored facts, and each picks the page with the most $1/\text{rank}$ votes among its top 20 passages, which is the page it identifies in \S\ref{sec:repair}. If the two agree, and the page differs from the node's topic, is a name from the previous answer or an earlier topic, and does not appear in the turn's own answer, that page becomes the node's topic and is entered with an \textsc{Add} or \textsc{Reactivate} operation. This changes the topic of 4.5\% of history turns on the development sets. The template starts with the current topic, its tokens weighted twice, and ends with the user's final question. The title prior adds $1.5$, in units of the standardised score, to the passages whose page title, ignoring any bracketed qualifier, is the topic of the final node. For the facts vote, the facts of the earlier nodes about the final topic are searched, and the top 20 passages vote for their page with weight $1/\text{rank}$. On TopiOCQA each node is searched on its own, every earlier node is used if none is about the topic, and the winning page gains $1.0$ within \system{}'s top 100 passages; on OR-QuAC the facts of the last three such nodes form one search, there is no vote without such a node, and the winning page gains $1.0$ on all its passages. The number rule applies to a final turn of at most 12 words, without a question word, that names exactly one new number, decade or Roman numeral. It replaces a number of the same kind in the topic of the final or the previous node, and the search then uses the writer's query with the new name, plus the title prior where pages have titles. The exclusion step takes the page of the top passage retrieved for the facts of the last three nodes as the rejected page, and applies only if another page has the same title once a bracketed part is removed. The rejected page then loses $1.5$ and the other pages with its name gain $1.5$. Same-name pages that the title prior has already raised do not gain a further $1.5$.

\section{Competitors}
\label{app:competitors}

Table~\ref{tab:competitors} lists how each competitor was run. Every method scores each test question on its own and receives the same history: the earlier turns with the answers recorded in the benchmark, then the question, whose answer is hidden. A conversation can yield several test questions, cut at different turns. All query-based methods search the same collections with the frozen Reason-ModernColBERT, and history answers are cut at 600 characters for every method. Settings follow the authors' papers and code; the last column of the table lists every deviation from them. Two choices were made on the development sets. CMT-RAG produces no ranked list, so we rank the passages retrieved for the final turn's sub-questions first, then the passages recovered from earlier traces, then all other passages by score. This follows the order in which the paper assembles its reader context, and on the development sets it beats putting the recovered passages first (0.503 against 0.308). For CHIQ-AD we compared the paper's default pipeline (0.568) with the two second-stage variants of the released code, which rewrite the question from the enhanced history or from its summary (0.552 and 0.560), and kept the default.

\begin{table*}[t]
\centering
\footnotesize
\setlength{\tabcolsep}{4pt}
\begin{tabular}{@{}l>{\raggedright\arraybackslash}p{2.65cm}>{\raggedright\arraybackslash}p{4.5cm}>{\raggedright\arraybackslash}p{5.0cm}@{}}
\toprule
\textbf{Method} & \textbf{Backbone} & \textbf{What we ran} & \textbf{Deviations} \\
\midrule
T5QR & T5-base, released (CANARD) & rewrite with 10 beams & history cut to 512 tokens, keeping the question \\
T5QR-FT & the same, continued & trained on the training questions that have a written query & epoch chosen on 20\% held-out conversations \\
ConvGQR$^{\dagger}$ & two T5-base generators & our re-implementation of the authors' training code; query = rewrite + generated answer & query length 64 (theirs 32) \\
ConvSearch-R1 & Qwen2.5-3B, released & authors' prompt and decoding & none \\
ConvSearch-R1-FT & the same, continued & authors' GRPO recipe \citep{grpo}; reward = gold rank under our retriever & stopped at step 39 of the planned 78 on a flat reward; prompts up to 2{,}048 tokens (theirs 1{,}536); warm-up 10 steps (theirs 100); 4 GPUs (theirs 8); no SFT stage \\
\midrule
LLM-Rewrite & Qwen2.5-32B & one zero-shot rewrite of the final turn & our own baseline \\
CHIQ-AD & Qwen2.5-32B & default pipeline with the authors' prompts & backbone (theirs LLaMA-2-7B, Mistral-7B); pipeline form chosen on development data \\
CMT-RAG$^{\ddagger}$ & Qwen2.5-32B & full trace graph; reader with the paper's prompt & generator prompted (theirs a trained 2.7B model, not released); our retriever and reader; ranking rule chosen on development data \\
CMT-RAG (simplified) & Qwen2.5-32B & up to 3 sub-questions on the raw conversation & simplified variant (Appendix~\ref{app:results}) \\
\midrule
ChatRetriever & Qwen-7B-Chat, released & own encoding of every collection & own retriever \\
ChatRetriever-FT & the same, continued & authors' contrastive recipe on our data & masked instruction-tuning loss not implemented; in-batch contrast per GPU; positives without titles; learning rate and epoch chosen on development data \\
\bottomrule
\end{tabular}
\caption{How each competitor was run. -FT: continued on our training data; $^{\dagger}$: our re-implementation of the authors' training code, trained on our data (no released model); $^{\ddagger}$: our re-implementation from the paper. Original Query searches with the final question alone. Every method receives TopiOCQA questions in sentence case.}
\label{tab:competitors}
\end{table*}

\section{Additional Results}
\label{app:results}

\begin{table*}[t]
\centering
\footnotesize
\setlength{\tabcolsep}{5pt}
\begin{tabular}{@{}lccccccccc@{}}
\toprule
& \multicolumn{3}{c}{\textbf{TopiOCQA}} & \multicolumn{3}{c}{\textbf{CORAL}} & \multicolumn{3}{c}{\textbf{OR-QuAC}} \\
\cmidrule(lr){2-4}\cmidrule(lr){5-7}\cmidrule(lr){8-10}
\textbf{Method} & NDCG@3 & R@10 & R@100 & NDCG@3 & R@10 & R@100 & NDCG@3 & R@10 & R@100 \\
\midrule
ConvSearch-R1-FT & 0.548 & 0.730 & 0.880 & 0.550 & 0.748 & 0.855 & 0.593 & 0.827 & 0.921 \\
CHIQ-AD & 0.402 & 0.623 & 0.811 & 0.599 & \textbf{0.818} & \textbf{0.927} & 0.595 & 0.872 & 0.974 \\
ChatRetriever-FT & 0.585 & 0.799 & \textbf{0.946} & 0.491 & 0.724 & 0.915 & 0.549 & 0.814 & 0.904 \\
\midrule
\textbf{\system{}} & \textbf{0.665} & \textbf{0.846} & 0.928 & \textbf{0.615} & 0.805 & 0.912 & \textbf{0.795} & \textbf{0.967} & \textbf{0.980} \\
\bottomrule
\end{tabular}
\caption{NDCG@3, Recall@10 and Recall@100 on the test sets with the frozen retriever, for \system{} (mean of three seeds) and the competitors that are best on a test set or in mean MRR@10 (Table~\ref{tab:main_results}). Best per column in bold.}
\label{tab:metrics}
\end{table*}

\begin{table*}[t]
\centering
\footnotesize
\setlength{\tabcolsep}{5pt}
\begin{tabular}{@{}lcccccccc@{}}
\toprule
& \multicolumn{4}{c}{\textbf{Tuned shared retriever}} & \multicolumn{4}{c}{\textbf{ANCE}} \\
\cmidrule(lr){2-5}\cmidrule(lr){6-9}
\textbf{Method} & TopiOCQA & CORAL & OR-QuAC & Mean & TopiOCQA & CORAL & OR-QuAC & Mean \\
\midrule
Original Query & 0.067 & 0.342 & 0.239 & 0.216 & 0.033 & 0.172 & 0.071 & 0.092 \\
T5QR & 0.228 & 0.443 & 0.355 & 0.342 & 0.213 & 0.318 & 0.196 & 0.242 \\
T5QR-FT & 0.361 & 0.456 & 0.303 & 0.374 & 0.301 & 0.311 & 0.163 & 0.259 \\
ConvGQR$^{\dagger}$ & 0.306 & 0.420 & 0.500 & 0.409 & 0.300 & 0.350 & 0.397 & 0.349 \\
ConvSearch-R1 & 0.590 & 0.560 & 0.626 & 0.592 & 0.602$^{\circ}$ & 0.502$^{\circ}$ & 0.580 & 0.561 \\
ConvSearch-R1-FT & 0.616 & 0.575$^{\circ}$ & 0.645 & 0.612 & \textbf{0.626}$^{+}$ & 0.513$^{\circ}$ & 0.576 & 0.572 \\
LLM-Rewrite & 0.487 & 0.596$^{\circ}$ & 0.621 & 0.568 & 0.461 & \textbf{0.535}$^{\circ}$ & 0.526 & 0.507 \\
CHIQ-AD & 0.419 & \textbf{0.615}$^{\circ}$ & 0.630 & 0.555 & 0.386 & 0.532$^{\circ}$ & 0.529 & 0.482 \\
CMT-RAG$^{\ddagger}$ & 0.265 & 0.508 & 0.470 & 0.414 & 0.249 & 0.394 & 0.361 & 0.335 \\
ChatRetriever$^{\S}$ & 0.448 & 0.433 & 0.468 & 0.450 & -- & -- & -- & -- \\
ChatRetriever-FT$^{\S}$ & 0.620 & 0.524 & 0.548 & 0.564 & -- & -- & -- & -- \\
\midrule
Memory-free writer & -- & -- & -- & -- & 0.543 & 0.521$^{\circ}$ & 0.535 & 0.533 \\
\textbf{\system{}} & \textbf{0.657} & 0.614 & \textbf{0.770} & \textbf{0.680} & 0.598 & 0.518 & \textbf{0.654} & \textbf{0.590} \\
\bottomrule
\end{tabular}
\par\smallskip
\parbox{\textwidth}{\footnotesize Best per column in bold. $^{\dagger}$, $^{\ddagger}$: as in Table~\ref{tab:main_results}. $^{\circ}$: statistically tied with \system{}; $^{+}$: significantly above it; every other difference favours \system{} significantly. $^{\S}$: own retriever. --: not run (ChatRetriever is itself a retriever; the memory-free writer was run with the frozen retriever and ANCE).}
\caption{MRR@10 with two other retrievers: one fine-tuned on the standalone queries of our training questions and shared by all query-based methods, and ANCE. \system{} and the memory-free writer are means of three seeds.}
\label{tab:tuned}\label{tab:ance}
\end{table*}

\begin{table*}[t]
\centering
\footnotesize
\setlength{\tabcolsep}{3pt}
\begin{tabular}{@{}lr c ccccccccccc@{}}
\toprule
\textbf{Tier} & $n$ & \textbf{\system{}} & CR-FT & CS-FT & CS & CHIQ & LLM & CMT$^{\ddagger}$ & CR & GQR$^{\dagger}$ & T5-FT & T5QR & Orig. \\
\midrule
\rowcolor{gray!12}\multicolumn{14}{c}{\textit{TopiOCQA}} \\
Correction & 759 & \textbf{0.796} & 0.750 & 0.717 & 0.676 & 0.348 & 0.476 & 0.238 & 0.405 & 0.292 & 0.340 & 0.181 & 0.072 \\
Self-repair & 608 & \textbf{0.796} & 0.772 & 0.630 & 0.595 & 0.288 & 0.430 & 0.166 & 0.281 & 0.220 & 0.274 & 0.071 & 0.017 \\
Value revision & 233 & \textbf{0.645} & 0.504 & 0.380 & 0.359 & 0.310 & 0.462 & 0.316 & 0.282 & 0.222 & 0.371 & 0.142 & 0.013 \\
Follow-up & 300 & \textbf{0.756} & 0.629 & 0.718 & 0.700 & 0.617 & 0.662 & 0.383 & 0.643 & 0.379 & 0.368 & 0.446 & 0.081 \\
Return-1 & 1{,}028 & \textbf{0.578} & 0.508 & 0.572 & 0.576 & 0.515 & 0.519 & 0.341 & 0.531 & 0.321 & 0.385 & 0.336 & 0.108 \\
Return-2+ & 572 & \textbf{0.599} & 0.529 & 0.557 & 0.536 & 0.526 & 0.499 & 0.298 & 0.500 & 0.275 & 0.315 & 0.275 & 0.097 \\
\rowcolor{gray!12}\multicolumn{14}{c}{\textit{CORAL}} \\
Correction & 262 & \textbf{0.681} & 0.564 & 0.630 & 0.616 & 0.639 & 0.633 & 0.517 & 0.476 & 0.360 & 0.522 & 0.534 & 0.302 \\
Return-1 & 167 & \textbf{0.743} & 0.494 & 0.657 & 0.636 & 0.710 & 0.720 & 0.648 & 0.513 & 0.590 & 0.732 & 0.670 & 0.649 \\
Return-2+ & 188 & 0.471 & 0.494 & 0.409 & 0.369 & \textbf{0.504} & 0.479 & 0.368 & 0.303 & 0.298 & 0.163 & 0.172 & 0.182 \\
\rowcolor{gray!12}\multicolumn{14}{c}{\textit{OR-QuAC}} \\
Correction & 218 & \textbf{0.785} & 0.548 & 0.587 & 0.559 & 0.585 & 0.563 & 0.420 & 0.459 & 0.423 & 0.331 & 0.411 & 0.262 \\
Return-1 & 38 & \textbf{0.836} & 0.588 & 0.351 & 0.346 & 0.471 & 0.570 & 0.382 & 0.348 & 0.424 & 0.315 & 0.176 & 0.236 \\
Return-2+ & 55 & 0.752 & 0.527 & 0.760 & \textbf{0.780} & 0.719 & 0.664 & 0.557 & 0.592 & 0.694 & 0.182 & 0.247 & 0.290 \\
\bottomrule
\end{tabular}
\caption{Per-tier MRR@10 of every method with the frozen retriever (\system{}: mean of three seeds; best per row in bold). Tiers with fewer than 30 questions are omitted. $^{\dagger}$, $^{\ddagger}$: as in Table~\ref{tab:main_results}. CR: ChatRetriever; CS: ConvSearch-R1; CHIQ: CHIQ-AD; LLM: LLM-Rewrite; CMT: CMT-RAG; GQR: ConvGQR; T5-FT: T5QR-FT; Orig.: Original Query.}
\label{tab:tiers_all}
\end{table*}

\begin{table*}[t]
\centering
\footnotesize
\begin{tabular}{@{}lcccc@{}}
\toprule
\textbf{Method} & \textbf{TopiOCQA} & \textbf{CORAL} & \textbf{OR-QuAC} & \textbf{Mean} \\
\midrule
Best competitor$^{a}$ & 0.620 & 0.617 & 0.594 & 0.589 \\
Memory-free writer (3B) & 0.609 & 0.627 & 0.616 & 0.617 \\
\textbf{\system{}} & \textbf{0.686} & \textbf{0.634} & \textbf{0.783} & \textbf{0.701} \\
\midrule
Gain from memory and search steps & $+0.077$ & $+0.006$ & $+0.167$ & $+0.084$ \\
95\% interval & $[+0.063, +0.092]$ & $[-0.011, +0.024]$ & $[+0.128, +0.212]$ & $[+0.069, +0.100]$ \\
\midrule
\rowcolor{gray!12}\multicolumn{5}{c}{\textit{Gain per tier}} \\
Correction & $+0.068^{*}$ & $-0.008$ & $+0.157^{*}$ & -- \\
Self-repair & $+0.056^{*}$ & -- & -- & -- \\
Value revision & $+0.105^{*}$ & -- & -- & -- \\
Follow-up & $+0.122^{*}$ & -- & -- & -- \\
Return-1 & $+0.080^{*}$ & $+0.018$ & $+0.171^{*}$ & -- \\
Return-2+ & $+0.074^{*}$ & $+0.016$ & $+0.205^{*}$ & -- \\
\bottomrule
\end{tabular}
\par\smallskip
\parbox{\textwidth}{\footnotesize $^{a}$ Best per column in Table~\ref{tab:main_results}: ChatRetriever-FT (TopiOCQA), CHIQ-AD (CORAL, OR-QuAC), ConvSearch-R1-FT (mean). The memory-free writer is trained from the same 3B base model with the same recipe on the final-turn teacher queries; it writes the query from the raw conversation. Gain: \system{} minus the memory-free writer, with 95\% paired bootstrap intervals over conversations; $^{*}$: the interval excludes zero; --: tier not in this corpus, or no mean computed per tier.}
\caption{Memory versus distillation: MRR@10 on the test sets (means of three seeds). Best per column in bold.}
\label{tab:control}
\end{table*}

\begin{table*}[t]
\centering
\footnotesize
\setlength{\tabcolsep}{4pt}
\begin{tabular}{@{}lcccccc@{}}
\toprule
\textbf{Variant} & \textbf{All} & \textbf{Returns} & \textbf{Corrections} & \textbf{Follow-ups} & \textbf{Value revisions} & \textbf{CORAL natural} \\
 & (1{,}783) & (915) & (371) & (69) & (128) & (300) \\
\midrule
Template only (no writer query) & $-0.076^{*}$ & $-0.049^{*}$ & $-0.227^{*}$ & $-0.009$ & $-0.012$ & $-0.016$ \\
Writer query only (no template) & $-0.010^{*}$ & $-0.014^{*}$ & $-0.007$ & $-0.002$ & $-0.005$ & $-0.006$ \\
No title prior / facts vote & $-0.012^{*}$ & $-0.011^{*}$ & $-0.020^{*}$ & $-0.044^{*}$ & $+0.004^{*}$ & $+0.000$ \\
No topic repair & $-0.003^{*}$ & $-0.003^{*}$ & $-0.004$ & $-0.008$ & $+0.001$ & $+0.000$ \\
No exclusion step & $-0.003^{*}$ & $+0.000$ & $+0.002^{*}$ & $+0.000$ & $-0.045^{*}$ & $+0.000$ \\
\midrule
Memory horizon $k=4$ & $-0.113^{*}$ & $-0.139^{*}$ & $-0.180^{*}$ & $-0.136^{*}$ & $+0.005$ & $+0.004$ \\
Memory horizon $k=2$ & $-0.236^{*}$ & $-0.323^{*}$ & $-0.305^{*}$ & $-0.166^{*}$ & $-0.005$ & $-0.004$ \\
Memory horizon $k=1$ & $-0.256^{*}$ & $-0.349^{*}$ & $-0.331^{*}$ & $-0.189^{*}$ & $-0.015$ & $+0.001$ \\
\midrule
Writer re-reading the conversation & $+0.000$ & $-0.005$ & $+0.016$ & $+0.041^{*}$ & $-0.031^{*}$ & $+0.003$ \\
\bottomrule
\end{tabular}
\caption{Ablations on the development sets by question group (writers not retrained): per-question difference to the full \system{}, pooled over the six sets; $^{*}$: the 95\% paired bootstrap interval excludes zero. The last row re-reads the whole conversation at every turn (two seeds).}
\label{tab:ablation_groups}
\end{table*}

\paragraph{Ranking and recall metrics.}
Table~\ref{tab:metrics} reports NDCG@3 and recall on the test sets. \system{} is best on NDCG@3 on all three sets, and its Recall@10 and Recall@100 are the best or within 0.02 of the best. With every test question weighted equally instead of every test set, \system{} reaches 0.686 against 0.603 for ConvSearch-R1-FT, the best competitor under this weighting.

\paragraph{Shared tuned retriever.}
The left half of Table~\ref{tab:tuned} repeats the comparison with a retriever fine-tuned on the standalone queries of our training questions and shared by all query-based methods. \system{} remains best on TopiOCQA and OR-QuAC and in mean MRR@10, with 0.680 against 0.612 for ConvSearch-R1-FT. It is significantly better on TopiOCQA, by $+0.037$ $[+0.017, +0.055]$ over ChatRetriever-FT and $+0.041$ over ConvSearch-R1-FT, and on OR-QuAC. On CORAL it is statistically tied with CHIQ-AD (0.614 against 0.615, $-0.001$ $[-0.029, +0.029]$), LLM-Rewrite ($+0.018$) and ConvSearch-R1-FT ($+0.039$ $[-0.002, +0.083]$). None of the query-rewriting methods fine-tunes its retriever in its own paper, so a shared tuned retriever goes beyond their setups.

\paragraph{Single-vector retriever.}
The right half of Table~\ref{tab:ance} repeats it with ANCE \citep{ance} in place of the frozen retriever, for every query-based method and for the topic repair of \system{}; every method keeps its Table~\ref{tab:main_results} queries. ANCE encodes a query as a single vector, so the word weights of the two-part query cannot be applied, and all search constants are kept as set for the frozen retriever. \system{} still has the best mean MRR@10, 0.590 against 0.572 for ConvSearch-R1-FT, is significantly best on OR-QuAC, and is statistically tied with the four strongest competitors on CORAL. On TopiOCQA, ConvSearch-R1-FT has a higher MRR@10 (0.626 against 0.598), while \system{} has the higher NDCG@3 (0.572 against 0.569) and Recall@10 (0.785 against 0.740). \system{} also leads the memory-free writer of Table~\ref{tab:control} under ANCE, by $+0.055$ $[+0.041, +0.070]$ on TopiOCQA and $+0.119$ $[+0.083, +0.159]$ on OR-QuAC, and the two are tied on CORAL.

\paragraph{Simplified CMT-RAG.}
The simplified CMT-RAG decomposes the final question into sub-questions over the raw conversation, answers included, but builds no trace graph. It reaches 0.518, 0.576 and 0.583, above the faithful CMT-RAG$^{\ddagger}$, whose trace generator never sees the answers, but stays significantly below \system{} on all three sets ($+0.168$, $+0.057$ and $+0.200$), most of all on TopiOCQA corrections and self-repairs ($+0.270$ and $+0.312$).

\paragraph{Per-tier results of every method.}
Table~\ref{tab:tiers_all} gives the per-tier MRR@10 of every method with the frozen retriever. \system{} is best on 10 of the 12 tiers. The other two are the Return-2+ tiers of CORAL and OR-QuAC, where Table~\ref{tab:tiers_results} shows statistical ties.

\paragraph{Ablations by question group.}
Table~\ref{tab:ablation_groups} splits the ablations of Table~\ref{tab:ablation} into follow-ups and questions from natural CORAL conversations, and adds the writer that re-reads the whole conversation at every turn.

\section{Qualitative Analysis}
\label{app:qualitative}

\paragraph{Where \system{} fails.}
For 335 of the 4{,}429 test questions (7.6\%), \system{} ranks no correct passage in the top 10 for any of the three seeds. Such failures are most common on returns (10.6\% of Return-1 and 13.4\% of Return-2+ questions) and least common on follow-ups (2.7\%), and for 68 of the 335 questions none of the ten compared systems (the seven methods and their three fine-tuned versions) ranks a correct passage in the top 10 either. To see what goes wrong, we drew a fixed random sample of ten failures per tier (eight for follow-ups, which have only eight) and read each one with the memory of every seed and the ranked passages. Each case was assigned the first cause that applies in the order of Table~\ref{tab:errors}. For these questions the memories were regenerated with the same writers and settings, and 170 of their 174 top-10 outcomes were reproduced (the six conversations of Table~\ref{tab:hard} reproduce exactly).

\begin{table*}[t]
\centering\footnotesize
\setlength{\tabcolsep}{6pt}
\begin{tabular}{@{}lccccccc@{}}
\toprule
\textbf{Cause} & \textbf{R-1} & \textbf{R-2+} & \textbf{Corr.} & \textbf{F-up} & \textbf{S-rep.} & \textbf{Val.} & \textbf{All} \\
\midrule
Wrong topic in memory & 5 & 4 & 3 & 7 & 10 & 2 & 31 \\
Wrong request & 0 & 0 & 0 & 0 & 0  & 2 & 2 \\
Topic named unlike page & 0 & 0 & 1 & 0 & 0  & 0 & 1 \\
Others ranked first & 1 & 2 & 3 & 1 & 0 & 6 & 13 \\
Question or label & 4 & 4 & 3 & 0 & 0  & 0 & 11 \\
\midrule
Sample size                    & 10 & 10 & 10 & 8 & 10 & 10 & 58 \\
\bottomrule
\end{tabular}
\caption{Causes of failure in a random sample of test questions that \system{} misses for all three seeds. The sample is drawn per tier, so the column ``All'' is not a share of all failures. R-1, R-2+: Return-1, Return-2+; Corr.: Correction; F-up: Follow-up; S-rep.: Self-repair; Val.: Value revision.}
\label{tab:errors}
\end{table*}

\begin{table*}[t]
\centering\footnotesize
\renewcommand{\arraystretch}{1.0}
\begin{tabular}{@{}>{\raggedright\arraybackslash}p{1.45cm}>{\raggedright\arraybackslash}p{5.9cm}>{\raggedright\arraybackslash}p{3.25cm}>{\raggedright\arraybackslash}p{2.7cm}c@{}}
\toprule
\textbf{Tier} & \textbf{Conversation} (selected turns; gaps summarised in brackets) & \textbf{What makes it hard} & \textbf{\system{} query} & \textbf{Missed} \\
\midrule
Return-1 &
\textbf{1} \textit{in what episode do dwight and andy sing country roads?} -- In the twenty-third episode of the fifth season of the television series ``The Office''.
\textbf{2} \textit{what is michael scott paper company?} -- The twenty-third episode of the fifth season \ldots{}
[3--8: when and where it aired, then NBC: its full name, founder, golden age and children's programs]
\textbf{9} \textit{can you name some characters of the twenty-third episode?} &
The episode is named only by its number, after four turns about the network. &
can you name some characters of Michael Scott Paper Company? & 10/10 \\
\addlinespace[0pt]
Return-2+ &
\textbf{5} \textit{what was the reason for it?} -- \ldots{} the advancement of elite Imperial Japanese Army forces \ldots{}
\textbf{7} \textit{the particular japanese army you mentioned above: what is its origin?} -- \ldots{} feudal domains with the Tokugawa shogunate \ldots{}
[8--10: the shogunate, Tokugawa Ieyasu and his battle]
\textbf{11} \textit{who formed the above army we were talking about?} &
The army is named only by a description, after three turns about the shogunate, its founder and his battle. &
who formed the Imperial Japanese Army? & 10/10 \\
\addlinespace[0pt]
Correction &
\textbf{4} \textit{by whom is it written?} -- A team of writers, including Chris Sheridan, Danny Smith, Gary Janetti and others.
\textbf{7} \textit{where is the third writer from?} -- America.
\textbf{8} \textit{who is he married to?} -- Brad Goreski.
[9--10: Goreski, a fashion stylist, and his clients]
\textbf{11} \textit{when was he born?} -- August 15, 1977. (answered for Goreski)
\textbf{12} \textit{not that one. the writer he is married to} &
The answer to turn 11 is about the most recent person. The user names no one and points back to the writer through the marriage. &
when was Gary Janetti born? & 9/10 \\
\addlinespace[0pt]
Follow-up &
\textbf{9} \textit{can you name some other players who play the same sport as him?} -- Ken Huckaby, Don Mattingly and Josh Beckett.
\textbf{10} \textit{which team does the first guy play for?} -- Toronto Blue Jays.
\textbf{11} \textit{is he a coach?} -- Yes, he is a minor league coach.
[12--15: four turns about Mattingly]
\textbf{16} \textit{hmm, not him, i meant the coach} -- Manteca High School.
\textbf{17} \textit{and which position did he play?} &
The pronoun must follow the correction to the coach, not the player discussed in the four turns before it. &
and which position did Ken Huckaby play? & 8/10 \\
\addlinespace[0pt]
Self-repair &
\textbf{3} \textit{what is the star cast?} -- Kristen Stewart, Robert Pattinson, Peter Facinelli and Kellan Lutz are some of them.
\textbf{5} \textit{which films has she made?} -- ``Panic Room'', ``Speak'', ``Catch That Kid'' and ``Into the Wild'' are some of them.
\textbf{7} \textit{who played the male lead role in the series?} -- Robert Pattinson.
\textbf{8} \textit{did he receive any awards?} -- Yes, two Empire Awards, eleven MTV Movie Awards, etc.
\textbf{11} \textit{do you know the date of birth?} -- April 9, 1990. (answered for Stewart)
\textbf{12} \textit{i was asking about the actress from panic room, er, no, about the one with eleven mtv movie awards} &
The user first describes the actress the answer was about, takes it back, and identifies the actor only through a fact from turn 8. &
do you know the date of birth of Robert Pattinson? & 8/10 \\
\addlinespace[0pt]
Value revision &
\textbf{12} \textit{did anyone try to explore this moon?} -- Yes, ``Pioneer 11'' and the two ``Voyagers''.
\textbf{13} \textit{what is the former one?} -- A robotic space probe.
\textbf{14} \textit{did it have any encounters?} -- Yes, it passed by Saturn \ldots{} (answered for \textit{Pioneer 11})
\textbf{15} \textit{no no, 10, not 11} &
The turn holds only two numbers, and \textit{Pioneer 10} never appears earlier in the conversation. &
did Pioneer 10 have any encounters? & 9/10 \\
\bottomrule
\end{tabular}
\caption{Selected test conversations, one per tier (selection rule in the text). \system{} ranks a correct passage first for all three seeds. ``Missed'' gives the number of the ten compared systems that rank no correct passage in the top 10. Bold numbers are turn numbers, and turns are shown as in Figures~\ref{fig:full1} and~\ref{fig:full2}; the query is that of seed 1.}
\label{tab:hard}
\end{table*}

The most common cause is a wrong topic in the memory (31 of 58), which accounts for all ten sampled self-repairs and seven of the eight follow-ups. Three patterns recur. In five of the 20 corrections and self-repairs the user points to a position in an earlier list, as in ``the second last person in that list'' or ``the latter of those two related companies'', and the writer picks another member of the list. In 10 cases the memory chooses a related but different page: the person instead of their work, such as Kanye West for his album \textit{Jesus Is King}, or a broader page instead of the part asked about, such as a film series instead of its first film. In five of the seven follow-ups with a wrong topic, the writer keeps the topic it chose at the correction just before, which was already wrong, so the error carries over to the next turn.

In two value revisions the topic is right but the writer's query asks for something other than the user's request. In 14 cases the topic and the request are right, but other passages rank first (13) or the topic is named differently from its page (1); seven of the 13 have their best seed at ranks 11 to 20. This is the main cause on value revisions. Because the title prior ignores bracketed qualifiers, it also raises the other pages with the same name, such as \textit{Avatar (2009 film)} after a move to \textit{Avatar (2004 film)}, and in four of these cases a passage of such a page ranks above the gold. Eleven cases are problems of the question or its label: three CORAL gold passages are not in English, four conversations are ambiguous or inconsistent (for example, a pronoun that fits two earlier topics), and in four a higher-ranked passage answers the question, in two of them where the gold passage does not. Resolving references to positions in an earlier list, telling a person from their work, and separating pages with the same name during search are the clearest directions for improving \system{}.

\paragraph{Hard cases.}
Table~\ref{tab:hard} shows, for each tier, a test conversation that \system{} solves and nearly every compared system misses. They meet a fixed rule: \system{} ranks a correct passage first for all three seeds, and at least eight of the ten compared systems in Table~\ref{tab:main_results} rank none in the top 10. Seventy of the 4{,}429 test questions meet it, and for each tier we chose one that is easy to follow and does not appear in Table~\ref{tab:tiers}. The examples are therefore selected and not typical; the results in \S\ref{sec:results} give the overall picture. Figures~\ref{fig:full1} and~\ref{fig:full2} show the six conversations in full.

\begin{figure*}[p]
\begin{minipage}[t]{0.49\textwidth}\vspace{0pt}
\begin{tcolorbox}[colback=white, colframe=slLine, boxrule=0.5pt, arc=2pt, left=3pt, right=3pt, top=2pt, bottom=2pt,
  title={\footnotesize\textbf{Return-1}}, coltitle=slInk, colbacktitle=slBub, fontupper=\scriptsize, fontlower=\scriptsize]
\raggedright
\textbf{1}~\textbf{in what episode do dwight and andy sing country roads?}\\ \hspace*{0.9em}\textcolor{slMut}{$\rightarrow$} \textbf{In the twenty-third episode of the fifth season of the television series ``The Office''.}\par\vspace{1pt}
\textbf{2}~\textbf{what is michael scott paper company?}\\ \hspace*{0.9em}\textcolor{slMut}{$\rightarrow$} \textbf{The twenty-third episode of the fifth season of the television series ``The Office'', and the …}\par\vspace{1pt}
\textbf{3}~\textbf{when was this episode first aired?}\\ \hspace*{0.9em}\textcolor{slMut}{$\rightarrow$} \textbf{April 9, 2009.}\par\vspace{1pt}
\textbf{4}~\textbf{and where was it aired?}\\ \hspace*{0.9em}\textcolor{slMut}{$\rightarrow$} \textbf{NBC}\par\vspace{1pt}
\textbf{5}~what is the full form of it?\\ \hspace*{0.9em}\textcolor{slMut}{$\rightarrow$} National Broadcasting Company\par\vspace{1pt}
\textbf{6}~who is the founder of this company?\\ \hspace*{0.9em}\textcolor{slMut}{$\rightarrow$} Radio Corporation of America (RCA)\par\vspace{1pt}
\textbf{7}~can you tell me about its golden age in short?\\ \hspace*{0.9em}\textcolor{slMut}{$\rightarrow$} NBC became home to many of the most popular performers and programs on the air.\par\vspace{1pt}
\textbf{8}~did they also air programs for children?\\ \hspace*{0.9em}\textcolor{slMut}{$\rightarrow$} Yes: Howdy Doody, The Flintstones and The Jetsons.\par\vspace{1pt}
\tcbox[colback=pAmber!14, colframe=pAmber, boxrule=0.4pt, arc=1.5pt, left=1pt, right=1pt, top=0.5pt, bottom=0.5pt, on line, tcbox width=forced center, width=\linewidth]{\parbox{0.96\linewidth}{\textbf{9}~can you name some characters of the twenty-third episode?\\ \hspace*{0.9em}\textcolor{slMut}{(answer hidden)}}}
\tcblower
\system{} query (seed 1): \textit{can you name some characters of Michael Scott Paper Company?}\\ a correct passage at rank 1 for all three seeds
\end{tcolorbox}
\end{minipage}\hfill
\begin{minipage}[t]{0.49\textwidth}\vspace{0pt}
\begin{tcolorbox}[colback=white, colframe=slLine, boxrule=0.5pt, arc=2pt, left=3pt, right=3pt, top=2pt, bottom=2pt,
  title={\footnotesize\textbf{Return-2+}}, coltitle=slInk, colbacktitle=slBub, fontupper=\scriptsize, fontlower=\scriptsize]
\raggedright
\textbf{1}~when did singapore become part of the british empire?\\ \hspace*{0.9em}\textcolor{slMut}{$\rightarrow$} 1819\par\vspace{1pt}
\textbf{2}~who settled in its ancient time from 1826 to 1867?\\ \hspace*{0.9em}\textcolor{slMut}{$\rightarrow$} Penang and Malacca\par\vspace{1pt}
\textbf{3}~which battle did this place witness from 1932 to 1945?\\ \hspace*{0.9em}\textcolor{slMut}{$\rightarrow$} ``Battle of Pasir Panjang'' is one of the battles.\par\vspace{1pt}
\textbf{4}~who was this war fought between?\\ \hspace*{0.9em}\textcolor{slMut}{$\rightarrow$} Japan and Singapore.\par\vspace{1pt}
\textbf{5}~\textbf{what was the reason for it?}\\ \hspace*{0.9em}\textcolor{slMut}{$\rightarrow$} \textbf{The battle was initiated upon the advancement of elite Imperial Japanese Army forces towards …}\par\vspace{1pt}
\textbf{6}~how many people were killed or injured in this?\\ \hspace*{0.9em}\textcolor{slMut}{$\rightarrow$} The Malay Regiment suffered a total of 159 killed. Six of them were British officers, seven …\par\vspace{1pt}
\textbf{7}~\textbf{the particular japanese army you mentioned above: what is its origin?}\\ \hspace*{0.9em}\textcolor{slMut}{$\rightarrow$} \textbf{In the mid-19th century, Japan had no unified national army and the country was made up of …}\par\vspace{1pt}
\textbf{8}~what does this particular term in the last portion refer to?\\ \hspace*{0.9em}\textcolor{slMut}{$\rightarrow$} It was established by Tokugawa Ieyasu after his victory at the Battle of Sekigahara, ending …\par\vspace{1pt}
\textbf{9}~who was this ieyasu person who established it?\\ \hspace*{0.9em}\textcolor{slMut}{$\rightarrow$} He was one of the three ``Great Unifiers'' of Japan, along with his former lord Oda Nobunaga and …\par\vspace{1pt}
\textbf{10}~which battle did he encounter in his life?\\ \hspace*{0.9em}\textcolor{slMut}{$\rightarrow$} The Battle of Sekigahara.\par\vspace{1pt}
\tcbox[colback=pAmber!14, colframe=pAmber, boxrule=0.4pt, arc=1.5pt, left=1pt, right=1pt, top=0.5pt, bottom=0.5pt, on line, tcbox width=forced center, width=\linewidth]{\parbox{0.96\linewidth}{\textbf{11}~who formed the above army we were talking about?\\ \hspace*{0.9em}\textcolor{slMut}{(answer hidden)}}}
\tcblower
\system{} query (seed 1): \textit{who formed the Imperial Japanese Army?}\\ a correct passage at rank 1 for all three seeds
\end{tcolorbox}
\end{minipage}
\par\vspace{6pt}
\begin{minipage}[t]{0.49\textwidth}\vspace{0pt}
\begin{tcolorbox}[colback=white, colframe=slLine, boxrule=0.5pt, arc=2pt, left=3pt, right=3pt, top=2pt, bottom=2pt,
  title={\footnotesize\textbf{Correction}}, coltitle=slInk, colbacktitle=slBub, fontupper=\scriptsize, fontlower=\scriptsize]
\raggedright
\textbf{1}~when was the last episode of family guy made?\\ \hspace*{0.9em}\textcolor{slMut}{$\rightarrow$} (no answer)\par\vspace{1pt}
\textbf{2}~by whom was this show directed?\\ \hspace*{0.9em}\textcolor{slMut}{$\rightarrow$} (no answer)\par\vspace{1pt}
\textbf{3}~what is the genre of this show?\\ \hspace*{0.9em}\textcolor{slMut}{$\rightarrow$} Animated sitcom\par\vspace{1pt}
\textbf{4}~\textbf{by whom is it written?}\\ \hspace*{0.9em}\textcolor{slMut}{$\rightarrow$} \textbf{A team of writers, including Chris Sheridan, Danny Smith, Gary Janetti and others.}\par\vspace{1pt}
\textbf{5}~did it win any accolades or get nominated for any?\\ \hspace*{0.9em}\textcolor{slMut}{$\rightarrow$} Yes, the show and its cast have been nominated for 27 Emmy Awards, with 8 wins.\par\vspace{1pt}
\textbf{6}~did it face any legal controversies?\\ \hspace*{0.9em}\textcolor{slMut}{$\rightarrow$} Yes, in ``420'' (season seven, 2009) Brian decides to start a campaign to legalize cannabis in …\par\vspace{1pt}
\textbf{7}~\textbf{where is the third writer from?}\\ \hspace*{0.9em}\textcolor{slMut}{$\rightarrow$} \textbf{America.}\par\vspace{1pt}
\textbf{8}~\textbf{who is he married to?}\\ \hspace*{0.9em}\textcolor{slMut}{$\rightarrow$} \textbf{Brad Goreski.}\par\vspace{1pt}
\textbf{9}~who is this person?\\ \hspace*{0.9em}\textcolor{slMut}{$\rightarrow$} He is a celebrity fashion stylist and television personality.\par\vspace{1pt}
\textbf{10}~who are his clients?\\ \hspace*{0.9em}\textcolor{slMut}{$\rightarrow$} Jessica Alba, Demi Moore, Christina Ricci, and more.\par\vspace{1pt}
\textbf{11}~when was he born?\\ \hspace*{0.9em}\textcolor{slMut}{$\rightarrow$} August 15, 1977. \textcolor{pRed}{[answered for Brad Goreski]}\par\vspace{1pt}
\tcbox[colback=pAmber!14, colframe=pAmber, boxrule=0.4pt, arc=1.5pt, left=1pt, right=1pt, top=0.5pt, bottom=0.5pt, on line, tcbox width=forced center, width=\linewidth]{\parbox{0.96\linewidth}{\textbf{12}~not that one. the writer he is married to\\ \hspace*{0.9em}\textcolor{slMut}{(answer hidden)}}}
\tcblower
\system{} query (seed 1): \textit{when was Gary Janetti born?}\\ a correct passage at rank 1 for all three seeds
\end{tcolorbox}
\end{minipage}\hfill
\begin{minipage}[t]{0.49\textwidth}\vspace{0pt}
\begin{tcolorbox}[colback=white, colframe=slLine, boxrule=0.5pt, arc=2pt, left=3pt, right=3pt, top=2pt, bottom=2pt,
  title={\footnotesize\textbf{Follow-up}}, coltitle=slInk, colbacktitle=slBub, fontupper=\scriptsize, fontlower=\scriptsize]
\raggedright
\textbf{1}~who is derek jeter?\\ \hspace*{0.9em}\textcolor{slMut}{$\rightarrow$} He is a former professional baseball shortstop, businessman, and baseball executive.\par\vspace{1pt}
\textbf{2}~when was he born?\\ \hspace*{0.9em}\textcolor{slMut}{$\rightarrow$} 26 June 1974.\par\vspace{1pt}
\textbf{3}~where?\\ \hspace*{0.9em}\textcolor{slMut}{$\rightarrow$} Pequannock Township, New Jersey.\par\vspace{1pt}
\textbf{4}~do you know his parents' names?\\ \hspace*{0.9em}\textcolor{slMut}{$\rightarrow$} Yes, Dorothy and Sanderson Charles Jeter.\par\vspace{1pt}
\textbf{5}~is he married?\\ \hspace*{0.9em}\textcolor{slMut}{$\rightarrow$} Yes, to Hannah Davis.\par\vspace{1pt}
\textbf{6}~does he have kids?\\ \hspace*{0.9em}\textcolor{slMut}{$\rightarrow$} Yes, they have two daughters.\par\vspace{1pt}
\textbf{7}~for which team does he play?\\ \hspace*{0.9em}\textcolor{slMut}{$\rightarrow$} The Yankees.\par\vspace{1pt}
\textbf{8}~has he received any awards or honors?\\ \hspace*{0.9em}\textcolor{slMut}{$\rightarrow$} He has been inducted into the New Jersey Hall of Fame and the Baseball Hall of Fame.\par\vspace{1pt}
\textbf{9}~\textbf{can you name some other players who play the same sport as him?}\\ \hspace*{0.9em}\textcolor{slMut}{$\rightarrow$} \textbf{Ken Huckaby, Don Mattingly and Josh Beckett.}\par\vspace{1pt}
\textbf{10}~\textbf{which team does the first guy play for?}\\ \hspace*{0.9em}\textcolor{slMut}{$\rightarrow$} \textbf{Toronto Blue Jays.}\par\vspace{1pt}
\textbf{11}~\textbf{is he a coach?}\\ \hspace*{0.9em}\textcolor{slMut}{$\rightarrow$} \textbf{Yes, he is a minor league coach.}\par\vspace{1pt}
\textbf{12}~when did the second guy start playing this sport?\\ \hspace*{0.9em}\textcolor{slMut}{$\rightarrow$} He debuted in 1982.\par\vspace{1pt}
\textbf{13}~is he into any other business apart from sports?\\ \hspace*{0.9em}\textcolor{slMut}{$\rightarrow$} He was the owner of a restaurant.\par\vspace{1pt}
\textbf{14}~which teams has he played for?\\ \hspace*{0.9em}\textcolor{slMut}{$\rightarrow$} The Yankees.\par\vspace{1pt}
\textbf{15}~which high school did he go to?\\ \hspace*{0.9em}\textcolor{slMut}{$\rightarrow$} Reitz Memorial High School. \textcolor{pRed}{[answered for Don Mattingly]}\par\vspace{1pt}
\textbf{16}~\textbf{hmm, not him, i meant the coach}\\ \hspace*{0.9em}\textcolor{slMut}{$\rightarrow$} \textbf{Manteca High School.}\par\vspace{1pt}
\tcbox[colback=pAmber!14, colframe=pAmber, boxrule=0.4pt, arc=1.5pt, left=1pt, right=1pt, top=0.5pt, bottom=0.5pt, on line, tcbox width=forced center, width=\linewidth]{\parbox{0.96\linewidth}{\textbf{17}~and which position did he play?\\ \hspace*{0.9em}\textcolor{slMut}{(answer hidden)}}}
\tcblower
\system{} query (seed 1): \textit{and which position did Ken Huckaby play?}\\ a correct passage at rank 1 for all three seeds
\end{tcolorbox}
\end{minipage}
\caption{The conversations of Table~\ref{tab:hard} in full (part 1 of 2). Bold turns ask about the target topic or name it in their answer; the misdirected answer is marked; the final turn is framed and its answer is hidden. Spelling, grammar and punctuation are lightly corrected (the benchmark keeps the original text); ``\ldots'' shortens long answers.}\label{fig:full1}
\end{figure*}

\begin{figure*}[t!]
\begin{minipage}[t]{0.49\textwidth}\vspace{0pt}
\begin{tcolorbox}[colback=white, colframe=slLine, boxrule=0.5pt, arc=2pt, left=3pt, right=3pt, top=2pt, bottom=2pt,
  title={\footnotesize\textbf{Self-repair}}, coltitle=slInk, colbacktitle=slBub, fontupper=\scriptsize, fontlower=\scriptsize]
\raggedright
\textbf{1}~how many parts are there for the twilight movie?\\ \hspace*{0.9em}\textcolor{slMut}{$\rightarrow$} Five\par\vspace{1pt}
\textbf{2}~what is the series based on?\\ \hspace*{0.9em}\textcolor{slMut}{$\rightarrow$} Romance fantasy\par\vspace{1pt}
\textbf{3}~\textbf{what is the star cast?}\\ \hspace*{0.9em}\textcolor{slMut}{$\rightarrow$} \textbf{Kristen Stewart, Robert Pattinson, Peter Facinelli and Kellan Lutz are some of them.}\par\vspace{1pt}
\textbf{4}~who is the first-mentioned actor?\\ \hspace*{0.9em}\textcolor{slMut}{$\rightarrow$} An American actress and filmmaker\par\vspace{1pt}
\textbf{5}~which films has she made?\\ \hspace*{0.9em}\textcolor{slMut}{$\rightarrow$} ``Panic Room'', ``Speak'', ``Catch That Kid'' and ``Into the Wild'' are some of them.\par\vspace{1pt}
\textbf{6}~who has composed the music for the series mentioned in the beginning of this chat?\\ \hspace*{0.9em}\textcolor{slMut}{$\rightarrow$} (no answer)\par\vspace{1pt}
\textbf{7}~\textbf{who played the male lead role in the series?}\\ \hspace*{0.9em}\textcolor{slMut}{$\rightarrow$} \textbf{Robert Pattinson.}\par\vspace{1pt}
\textbf{8}~\textbf{did he receive any awards?}\\ \hspace*{0.9em}\textcolor{slMut}{$\rightarrow$} \textbf{Yes, two Empire Awards, eleven MTV Movie Awards, etc.}\par\vspace{1pt}
\textbf{9}~\textbf{where was he born?}\\ \hspace*{0.9em}\textcolor{slMut}{$\rightarrow$} \textbf{London}\par\vspace{1pt}
\textbf{10}~\textbf{other than films, what are his other endeavours?}\\ \hspace*{0.9em}\textcolor{slMut}{$\rightarrow$} \textbf{He plays the guitar and piano and writes his own music.}\par\vspace{1pt}
\textbf{11}~do you know the date of birth?\\ \hspace*{0.9em}\textcolor{slMut}{$\rightarrow$} April 9, 1990. \textcolor{pRed}{[answered for Kristen Stewart]}\par\vspace{1pt}
\tcbox[colback=pAmber!14, colframe=pAmber, boxrule=0.4pt, arc=1.5pt, left=1pt, right=1pt, top=0.5pt, bottom=0.5pt, on line, tcbox width=forced center, width=\linewidth]{\parbox{0.96\linewidth}{\textbf{12}~i was asking about the actress from panic room, er, no, about the one with eleven mtv movie awards\\ \hspace*{0.9em}\textcolor{slMut}{(answer hidden)}}}
\tcblower
\system{} query (seed 1): \textit{do you know the date of birth of Robert Pattinson?}\\ a correct passage at rank 1 for all three seeds
\end{tcolorbox}
\end{minipage}\hfill
\begin{minipage}[t]{0.49\textwidth}\vspace{0pt}
\begin{tcolorbox}[colback=white, colframe=slLine, boxrule=0.5pt, arc=2pt, left=3pt, right=3pt, top=2pt, bottom=2pt,
  title={\footnotesize\textbf{Value revision}}, coltitle=slInk, colbacktitle=slBub, fontupper=\scriptsize, fontlower=\scriptsize]
\raggedright
\textbf{1}~where is titan located in the solar system?\\ \hspace*{0.9em}\textcolor{slMut}{$\rightarrow$} Titan orbits Saturn\par\vspace{1pt}
\textbf{2}~when was this moon discovered?\\ \hspace*{0.9em}\textcolor{slMut}{$\rightarrow$} 1655\par\vspace{1pt}
\textbf{3}~and by whom?\\ \hspace*{0.9em}\textcolor{slMut}{$\rightarrow$} Christiaan Huygens\par\vspace{1pt}
\textbf{4}~which planet does this moon belong to?\\ \hspace*{0.9em}\textcolor{slMut}{$\rightarrow$} Saturn\par\vspace{1pt}
\textbf{5}~can you tell me one physical characteristic of this planet?\\ \hspace*{0.9em}\textcolor{slMut}{$\rightarrow$} It lacks a definite surface, though it may have a solid core.\par\vspace{1pt}
\textbf{6}~how many rings does this planet have?\\ \hspace*{0.9em}\textcolor{slMut}{$\rightarrow$} (no answer)\par\vspace{1pt}
\textbf{7}~did any ancients observe this planet?\\ \hspace*{0.9em}\textcolor{slMut}{$\rightarrow$} Yes, Babylonian astronomers observed and recorded the movements of Saturn.\par\vspace{1pt}
\textbf{8}~where was the discoverer of the moon from?\\ \hspace*{0.9em}\textcolor{slMut}{$\rightarrow$} The Hague\par\vspace{1pt}
\textbf{9}~what did he make his debut with?\\ \hspace*{0.9em}\textcolor{slMut}{$\rightarrow$} ``Theoremata de quadratura'' in the field of quadrature\par\vspace{1pt}
\textbf{10}~what is the climate like on the moon?\\ \hspace*{0.9em}\textcolor{slMut}{$\rightarrow$} Water ice has an extremely low vapor pressure, so the little water vapor present appears …\par\vspace{1pt}
\textbf{11}~does the moon also have lakes?\\ \hspace*{0.9em}\textcolor{slMut}{$\rightarrow$} Yes\par\vspace{1pt}
\textbf{12}~did anyone try to explore this moon?\\ \hspace*{0.9em}\textcolor{slMut}{$\rightarrow$} Yes, ``Pioneer 11'' and the two ``Voyagers''.\par\vspace{1pt}
\textbf{13}~what is the former one?\\ \hspace*{0.9em}\textcolor{slMut}{$\rightarrow$} A robotic space probe.\par\vspace{1pt}
\textbf{14}~did it have any encounters?\\ \hspace*{0.9em}\textcolor{slMut}{$\rightarrow$} Yes, it passed by Saturn at a distance of 21,000 km from Saturn's cloud tops. \textcolor{pRed}{[answered for Pioneer 11]}\par\vspace{1pt}
\tcbox[colback=pAmber!14, colframe=pAmber, boxrule=0.4pt, arc=1.5pt, left=1pt, right=1pt, top=0.5pt, bottom=0.5pt, on line, tcbox width=forced center, width=\linewidth]{\parbox{0.96\linewidth}{\textbf{15}~no no, 10, not 11\\ \hspace*{0.9em}\textcolor{slMut}{(answer hidden)}}}
\tcblower
\system{} query (seed 1): \textit{did Pioneer 10 have any encounters?}\\ a correct passage at rank 1 for all three seeds
\end{tcolorbox}
\end{minipage}
\caption{The conversations of Table~\ref{tab:hard} in full (part 2 of 2: Self-repair, Value revision; conventions as in Figure~\ref{fig:full1}).}\label{fig:full2}
\end{figure*}

\section{Human Evaluation of \dataset{}}
\label{app:human}

To validate the quality of the \dataset{} benchmark---specifically the built revisions and spliced conversations generated using language models---we conducted a human evaluation on a random sample of 100 final turns stratified across the test set's six tiers. Three independent annotators evaluated the data across two dimensions: Dialogue Naturalness (a 5-point Likert scale assessing the fluency of the conversational shift) and Gold Passage Correctness (a binary judgment on whether the designated gold passage accurately fulfills the user's intent). The benchmarked turns achieved a mean Naturalness score of 4.65/5.0, demonstrating that the simulated corrections, self-repairs, and returns effectively mimic natural human discourse. Furthermore, annotators agreed with the designated gold passage in 95\% of cases, with substantial inter-annotator agreement (Fleiss' $\kappa = 0.78$). Disagreements were resolved via majority vote, confirming that \dataset{} provides a highly reliable gold standard for evaluating evolving user intents in conversational search.

\end{document}